\documentclass{article}
\usepackage{amsfonts}
\usepackage{amssymb}
\usepackage{amsmath}
\usepackage{graphicx}
\usepackage{epstopdf}
\usepackage{slashed}
\usepackage{setspace}

\newcommand{\cl}{{\cal L}}

\newcommand{\be}{\begin{equation}}
\newcommand{\ee}{\end{equation}}

\def\be{\begin{equation}}
\def\ee{\end{equation}}
\def\bea{\begin{eqnarray}}
\def\eea{\end{eqnarray}}
\def\ba{\begin{array}}
\def\ea{\end{array}}
\def\bd{\begin{displaymath}}
\def\ed{\end{displaymath}}

\def\a{\alpha}
\def\b{\beta}

\def\d{\delta}
\def\e{\epsilon}           

\def\l{\lambda}
\def\m{\mu}
\def\n{\nu}
  
\def\s{\sigma}                                   

\def\D{\Delta}

\def\G{\Gamma}

\def\L{\Lambda}

\def\pa{\partial}                              
\def\>{\rangle} 
\def\<{\langle} 
\def\Dsl{D \hskip-.6em \raise1pt\hbox{$ / $ } }
\def\to{\rightarrow}

\def\pa{\partial}

\def\lab{\label}

\def \vy{{\vec y}}

\begin{document}

\vbox{\baselineskip14pt
} {~~~~~~~~~~~~~~~~~~~~~~~~~~~~~~~~~~~~
~~~~~~~~~~~~~~~~~~~~~~~~~~~~~~~~~~~
~~~~~~~~~~~~~~~~~ \footnotesize{SU-ITP-15/14, MIT-CTP/4598, MCTP-15-21}}

\begin{center}
\huge{AdS/CFT and the Little Hierarchy Problem}
\end{center}

\begin{center}
{\fontsize{13}{30}\selectfont Xi Dong$^{1,2}$, Daniel Z. Freedman$^{1,3}$, and Yue Zhao$^{1,4}$}
\end{center}

\begin{center}
\vskip 8pt
\textsl{ $^1$Stanford Institute for Theoretical Physics, Department of Physics, Stanford University,\\ Stanford, CA 94305, USA}

\vskip 7pt
\textsl{ $^2$School of Natural Sciences, Institute for Advanced Study, Princeton, NJ 08540, USA}

\vskip 7pt
\textsl{ $^3$Center for Theoretical Physics and Department of Mathematics, Massachusetts Institute of Technology, Cambridge, MA 02139, USA}

\vskip 7pt
\textsl{ $^4$Randall Laboratory of Physics, Department of Physics, University of Michigan, Ann Arbor, MI 48109, USA}
\end{center}

\abstract{ The AdS/CFT correspondence is applied to a close analogue of the little hierarchy problem in $AdS_{d+1}$, $d \geq 3$.  The new mechanism requires a Maxwell field that gauges a $U(1)_R$ symmetry in a bulk supergravity theory with a negative cosmological constant.  Supersymmetry is explicitly broken by a non-local boundary term with dimensionless coupling $h$.  Non-locality appears to cause no pathology, and the SUSY breaking deformation engendered is exactly marginal.  SUSY breaking effects in the bulk are computed using the $U(1)$ Ward identity.  Conformal dimensions and thus masses of scalar and spinor partners are split simply because they carry different R-charges.  However SUSY breaking effects cancel to all orders for R-neutral fields, even in diagrams with internal R-charged loops.  SUSY breaking corrections can be summed to all orders in $h$.  Diagrams involving graviton loops do not introduce any further SUSY breaking corrections.  A possible scenario for a flat spacetime limit is outlined. }

\section{Introduction}\lab{intro}

In a quantum field theory, scalar masses are naturally heavy due to quantum corrections  if they are not protected by a
symmetry. Global supersymmetry eliminates the most acute
quadratic sensitivity to unknown UV physics.  But supersymmetry is
usually assumed to be spontaneously broken and scalar masses
generically rise to the SUSY breaking scale.  LHC results
indicate that this scale  is larger than a few TeV.
The separation between the SUSY breaking
scale and the Higgs mass is the little hierarchy problem. A mechanism is
then needed to protect the Higgs mass.

There has already been much effort in model building  to address
the little hierarchy problem. For more detail,
see \cite{Craig:2013cxa,Dimopoulos:2014aua,Dimopoulos:2014psa} and
references therein. An especially interesting
cosmological relaxation mechanism \cite{Graham:2015cka} has recently been proposed.
This may offer a new direction to approach the little
hierarchy problem.

In this paper, we propose a mechanism to solve this problem in
$AdS_{d+1},~ d\ge 3$ based on the AdS/CFT correspondence. It
generalizes the $AdS_3$ model discussed in
\cite{Dong:2014tsa}. In that model we considered Chern-Simons gauge
fields $A_\mu,~\tilde A_\m$ which gauge a $U(1)_R\times \tilde
U(1)_R$ symmetry of a $D=3$ supergravity theory. The gauge fields
are bulk duals of conserved currents $J_i,\, \tilde J_i$ in the
boundary CFT. \, Supersymmetry is broken explicitly by a local
boundary term quadratic in the gauge fields, and these
fields propagate into the bulk and couple to R-charged matter
fields.  The Chern-Simons equation of motion is $F_{\m\n}=0$,  so
the bulk-to-boundary propagator is a total bulk derivative, i.e.
$K_{\mu i}(x,\vec w) = \pa_\mu\L_i(x,\vec w)$. Bulk integrals
induced by insertion of the SUSY breaking operator can then be
evaluated by partial integration and the $U( 1)$ Ward identity. The
SUSY breaking correction to any Witten diagram is then expressed by
a quite simple boundary integral.  The major results
are:

i. The scaling dimensions of CFT operators $O_c, \Psi_c$ which are superpartners are split because superpartners carry different R-charges.  AdS/CFT then implies that the masses of their dual bulk fields $\phi_c,~\psi_c$ are also split.

ii.  SUSY breaking corrections cancel completely for moduli fields which are $R$-neutral and massless in $AdS_3$, so moduli remain massless to all orders.

iii. Coupling constant relations required by SUSY are also corrected by this breaking mechanism.

Properties of the $AdS_3$ model are reviewed in Secs. 2 and 3 below, and we emphasize several features that are modified in the higher dimensional generalization  discussed in Sec. 4. It turns out that a successful extension can be constructed with a single Maxwell field $A_\mu$  which gauges a $U(1)_R$ symmetry of bulk supergravity.  The bulk-to-boundary propagator of a Maxwell field in  $AdS_{d+1}$ for $d\ge 3$ is not  a total derivative. But we devise a  SUSY breaking boundary term  that converts the
bulk-to-boundary propagator to a total derivative in the Witten diagrams that describe SUSY breaking corrections.  The SUSY breaking term is non-local but this appears to cause no pathology.  The features i.-iii. above hold in higher dimensions.  The possibility of a flat spacetime limit in which SUSY breaking effects survive is discussed in Sec. 5. Section 6 contains a summary and future outlook for theories of this type.  An appendix is devoted to the question of contact terms in the the divergence of the current-current correlator.\footnote{Some readers may be able to bypass the systematic review of the $AdS_3$ model and start instead in Sec. \ref{sec:JOcOc} or Sec. \ref{AdS4}.}

\section{Review of $AdS_3$}\lab{AdS3}
In this section, we review the features of the  $AdS_3$ model
in \cite{Dong:2014tsa}. This will serve to introduce the underlying
ideas and to motivate the changes needed for a viable model in
$AdS_{d+1}$.

Let us consider a supergravity theory in $D=3$ dimensions with
gauged R-symmetry group $U(1)\times \tilde{U}(1)$.  The R-gauge
bosons $A$ and $\tilde{A}$ have Chern-Simons dynamics. Gauging
requires a negative cosmological  constant, so the natural classical
solution is $AdS_3$. Pure supergravity models of this type were
first constructed by Achucarro and Townsend in 1986
\cite{Achucarro:1987vz} and general matter field couplings studied
in \cite{deWit:2003ja, deWit:2004yr,Deger:1999st}.

The bulk Chern-Simons action together with boundary terms that ensure a consistent variational problem is written as
\begin{eqnarray}
  \label{eq:CSterm}
S_{CS}=\int_{bulk}[\frac{k}{8\pi}A\wedge dA-\frac{\tilde
k}{8\pi}\tilde{A}\wedge
d\tilde{A}]-\int_{bdy}[\frac{ik}{16\pi}A\wedge * A+\frac{i\tilde
k}{16\pi}\tilde{A}\wedge*\tilde{A}] \,.
\end{eqnarray}
The boundary action chosen enforces that the anti-holomorphic
component of $A$ and the holomorphic component of $\tilde{A}$ vanish
on the boundary.

Now let us add  supersymmetric  matter fields to the theory. First,
we introduce the chiral multiplet $\Phi_m=(\phi_m, \psi_m,
\tilde\psi_m,...)$. The R-charges of the scalar $\phi_m$ vanish,
while its spinor partners $\psi_m$ and $\tilde\psi_m$ have R-charges
$(-1,0)$ and $(0,-1)$ respectively. The subscript $m$ stands for
``modulus" which means a scalar field with zero mass and vanishing
potential interactions.
We also add the chiral multiplet $\Phi_c=(\phi_c,
\psi_c,\tilde\psi_c,...)$ whose scalar and spinor components have
non-vanishing R-charges $(q,\tilde q)$ and  $(q-1,\tilde q)$,
$(q,\tilde q-1)$.
The subscript $c$ stands for ``charged".  The basic gauge invariant kinetic Lagrangian for $\phi_c$ is
\begin{equation}
  \label{eq:GaugeCoupling}
\cl_{\rm kin} =  \sqrt{g}\ D^\mu \phi_c^\dag D_\mu \phi_c =
\sqrt{g}\ g^{\mu\nu}(\partial_\mu-i q A_\mu-i\tilde{q}
\tilde{A}_\mu)\phi_c^\dag\ (\partial_\nu+i q A_\nu+i\tilde{q}
\tilde{A}_\nu)\phi_c\,.
\end{equation}
Non-trivial interactions between $\Phi_m$ and $\Phi_c$ can be
included. For example, the two-derivative cubic
coupling
\begin{eqnarray}
  \label{eq:cubic}
S_{cubic}=\int d^3x \sqrt{g}(\lambda\ \phi_m D^\mu\phi_c^\dag D_\mu\phi_c+h.c.)
\end{eqnarray}
arises from the K\"ahler $\s$-model of the theory.\footnote{A non-derivative cubic coupling is forbidden because the superpotential from which it arises  does not conserve R-charge. Results in the following sections remain valid for couplings with any even number of gauge covariant derivatives.}
Since $\phi_m$ is R-neutral, it does not couple to the R-gauge
bosons directly. However, it communicates  to $A$ and $\tilde{A}$ through
 loop diagrams involving R-charged particles such as $\phi_c$.

In this model, we introduce a special SUSY breaking mechanism. Explicit
SUSY breaking is generated by the  $AdS_3$ boundary term
\begin{eqnarray}
  \label{eq:SUSYBreak}
S_{SB}=\frac{h}{2}\int_{bdy}A\wedge \tilde{A} \,.
\end{eqnarray}
One can show that this SUSY breaking term is exactly marginal. As we review below,  this term breaks the
SUSY relation between boson and fermion mass spectra of R-charged multiplets as well as relations among
coupling constants. However, the masses of moduli fields vanish to all orders in the coupling $h$.

Since SUSY breaking term is exactly marginal, the AdS/CFT dual boundary theory is a $CFT_2$.  Indeed the dual of the SUSY breaking term (\ref{eq:SUSYBreak}) is
\begin{eqnarray}\label{sbterm3}
\label{eq:SUSYBreakCFT2} S_{SB,CFT}=h\int d^2 z J(z)\tilde{J}(\bar
z) \,,
\end{eqnarray}
where $J(z)$ and $\tilde J(\bar z)$ are holomorphic and
anti-holomorphic currents dual to $A$ and $\tilde{A}$
respectively. Therefore we can verify bulk calculations of SUSY breaking effects using well-known CFT methods.
      For example,
one can bosonize the currents as $J(z)\sim\partial_z
\eta(z)$ and $\tilde J(\bar z)\sim\partial_{\bar z} \tilde\eta(\bar
z)$. Exact marginality also follows from CFT arguments \cite{Chaudhuri:1988qb}.

Here we point out an important subtlety in order to avoid potential
confusions when we present the higher-dimensional generalization in
later sections. If the bosonized expressions for  $J$ and $\tilde{J}$ are inserted
into (\ref{eq:SUSYBreakCFT2}), it appears that one can integrate by
parts and use current conservation to remove
the SUSY breaking term. However, currents are conserved only when equations of motion are satisfied,  and we cannot naively use
equations of motion in a CFT Lagrangian. The observables in a CFT are
correlation functions, not S-matrices, as there is no
well-defined asymptotic free particle state.  Correlation functions, unlike S-matrix elements, can be sensitive to field redefinitions.  Thus the SUSY breaking term (\ref{eq:SUSYBreakCFT2}) cannot be trivialized or removed using
equations of motion. The generalization to higher-dimensional $AdS$
space has a similar feature.  For both $AdS_3$ (Sec. 4 of \cite{Dong:2014tsa}) and $AdS_{d+1}$ (Sec. 4.4.2 below) we have used conformal perturbation theory to calculate the shift of scaling dimension.  This shows that the deformation (\ref{sbterm3}) and its higher-dimensional generalization (\ref{eq:SUSYBreakCFTHigh}) are nontrivial.

\subsection{Mass Spectra} In this section, we explicitly show how the
mass spectra are affected by turning on the SUSY breaking term in
Eq. (\ref{eq:SUSYBreak}). Using the $AdS/CFT$ dictionary, studying the change of the mass of a particle is equivalent to studying the shift of
conformal dimension of the dual CFT operator. In $AdS_3$, they are
related by
\begin{eqnarray}
  \label{eq:MassDim}
\Delta_B&=&1+\sqrt{1+(m_B L)^2} \,,\nonumber\\
\Delta_F&=&1+|m_F L| \,.
\end{eqnarray}
When SUSY is a good symmetry, one has
\begin{eqnarray}
  \label{eq:SUSYMatch}
\Delta_F=\Delta_B+\frac{1}{2} \,.
\end{eqnarray}
A violation of this relation is a definite signal of SUSY
breaking. In this section, we show that after we turn on the SUSY
breaking term, i.e. Eq. (\ref{eq:SUSYBreak}), the relation
(\ref{eq:SUSYMatch}) is no  longer preserved.

In Sec. \ref{sec:R-chargedMass}, we show that the mass spectra are
changed in a non-supersymmetric way. In Sec.
\ref{sec:R-neutralMass}, we show that the mass, or even the full
self-potential, of the moduli field remains zero at all loop order.

\subsubsection{R-charged particle}\label{sec:R-chargedMass}

\begin{figure}[h]
\begin{center}
\includegraphics[width=0.2\textwidth]{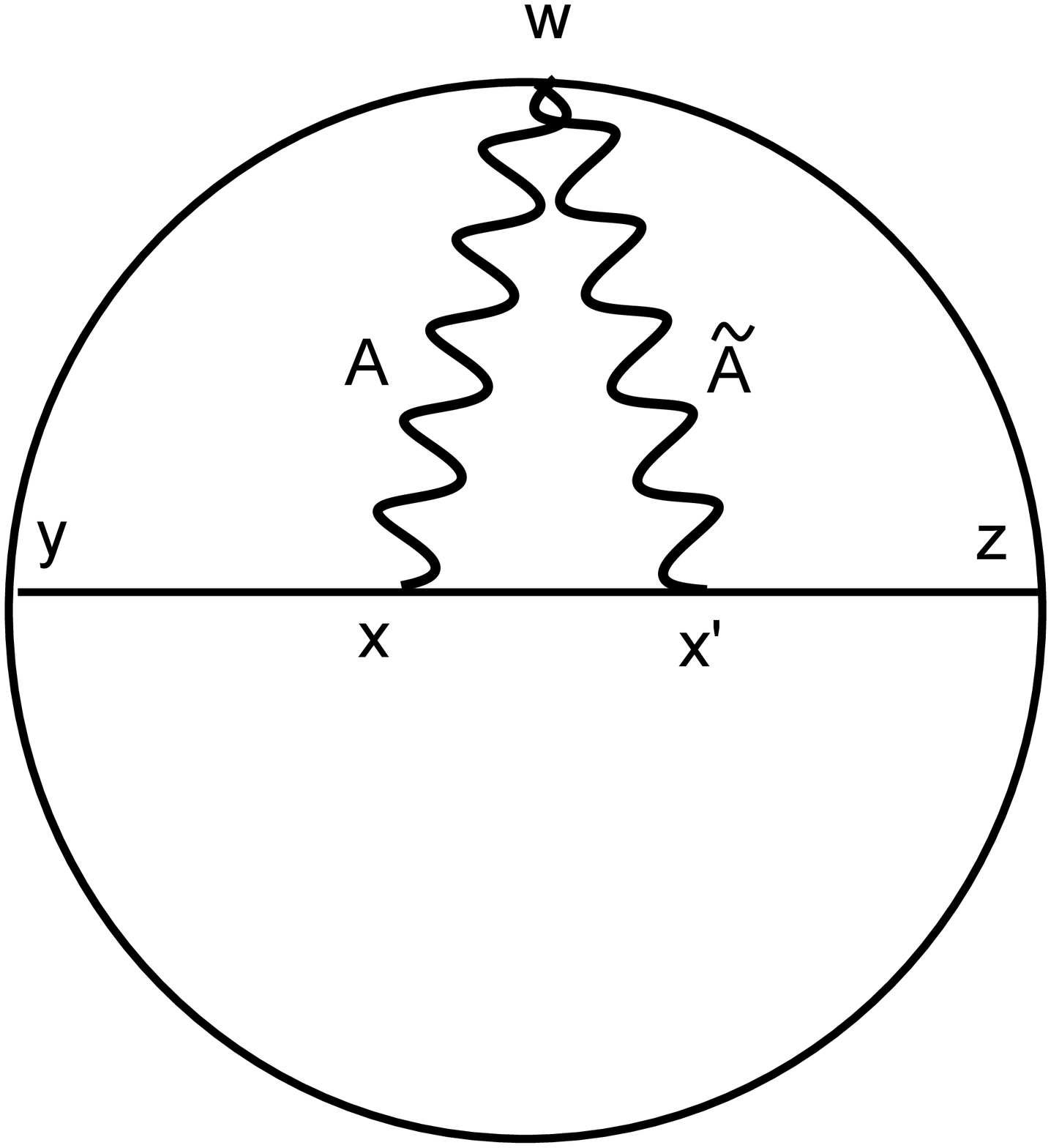}
\hspace*{0.35cm}
\includegraphics[width=0.2\textwidth]{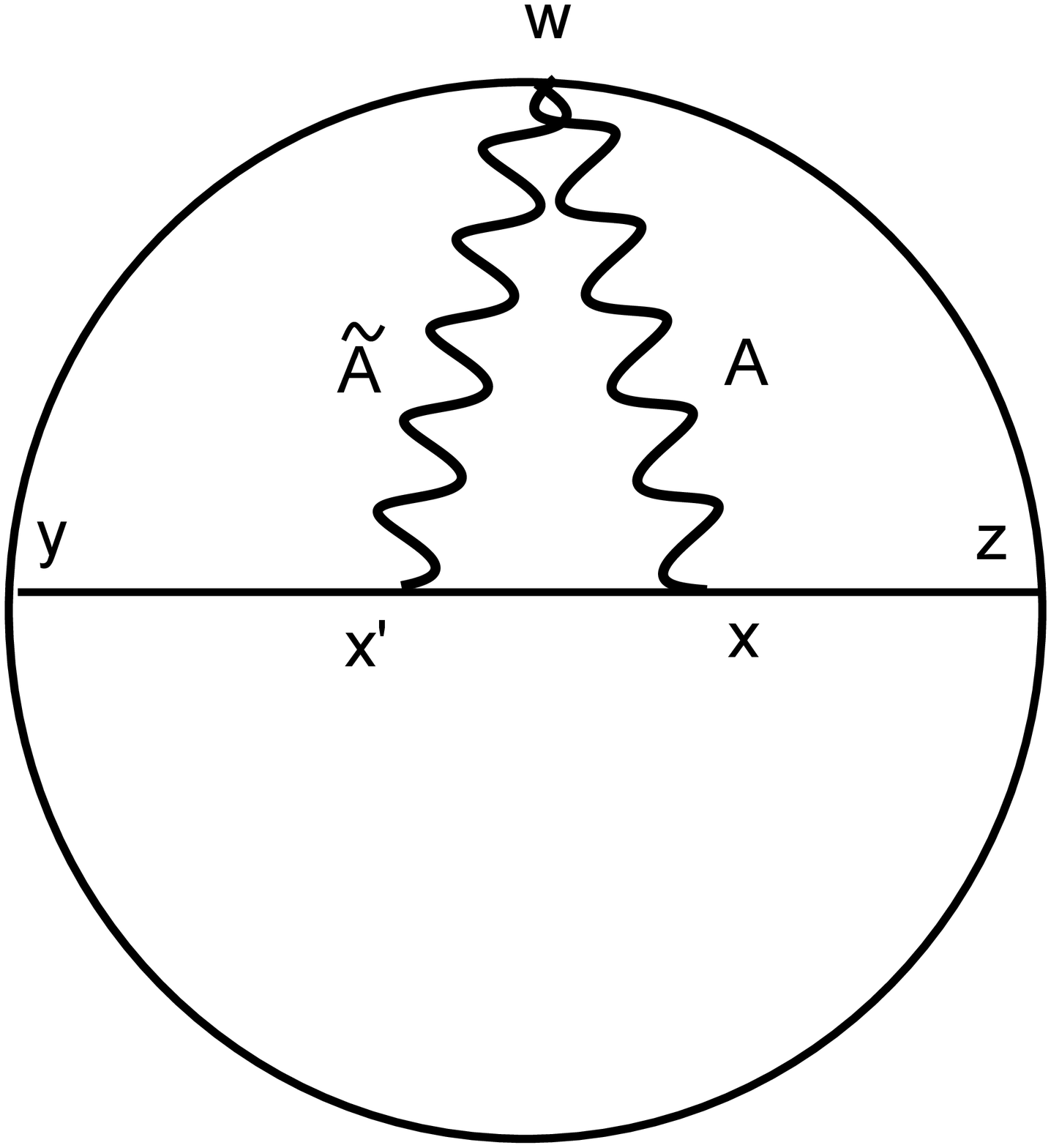}
\hspace*{0.35cm}
\includegraphics[width=0.2\textwidth]{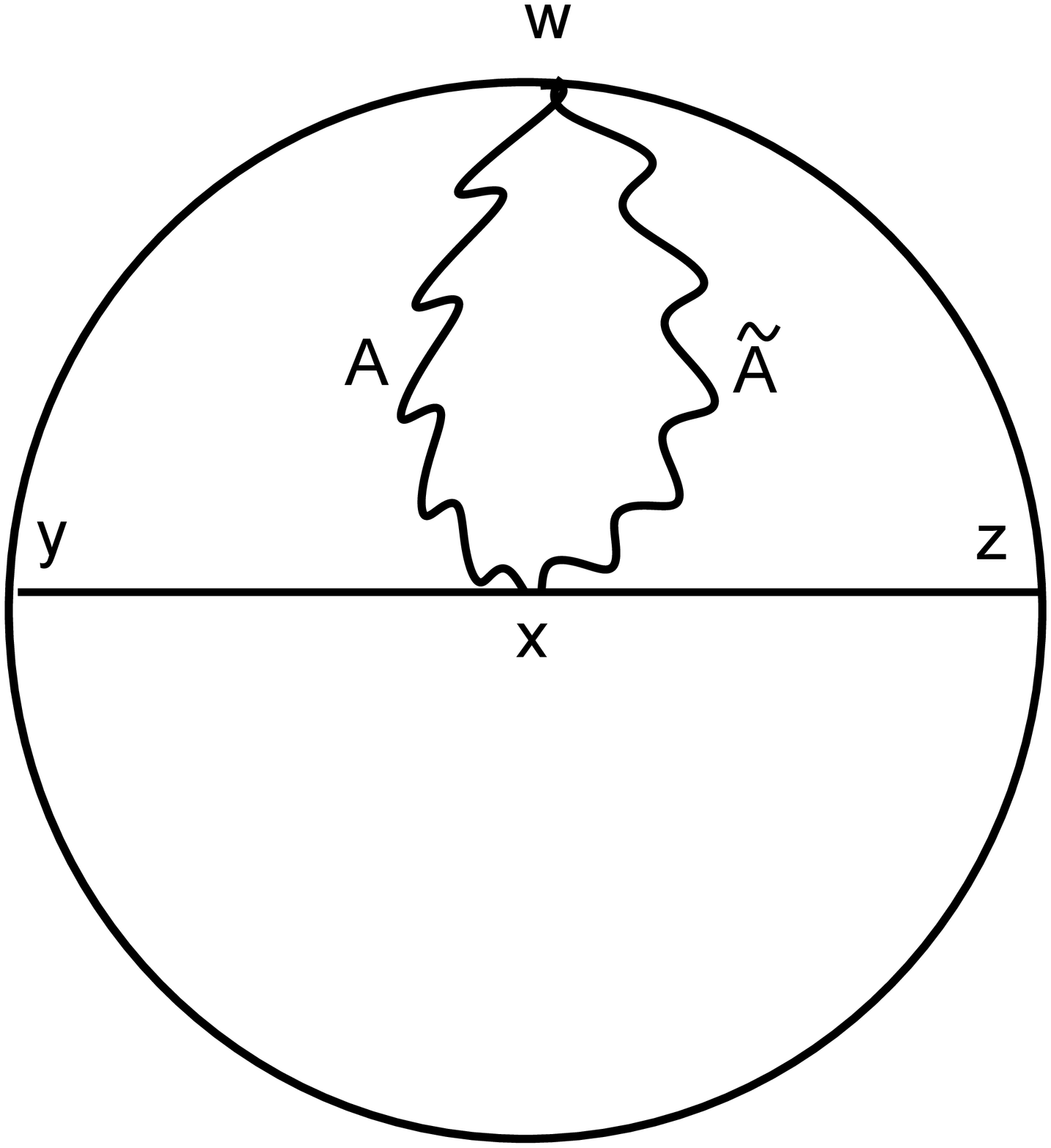}
\caption{The relevant diagrams for the leading order mass
deformation of a charged scalar field.} \label{fig:Charged}
\end{center}
\end{figure}

Let us start with the leading order mass correction for a bulk scalar $\phi_c$ whose
R-charges are $(q,\tilde{q})$. Because it is charged, this field couples directly
to $A_\mu,~\tilde A_\m$  through the gauge vertices in (\ref{eq:GaugeCoupling}).
To first order in $h$, the shift of  mass or  conformal dimension can be calculated by studying the insertion of $S_{\rm SB}$ in the  two-point
correlation function $\langle O_c^\dag(\vec{y})
O_c(\vec{z})\rangle$, where $O_c$ is the CFT dual
to $\phi_c$.
The relevant Witten diagrams are shown in Fig.
\ref{fig:Charged}.

The essential ingredients in the calculation of these diagrams are the bulk-to-boundary propagator of the Chern-Simons gauge field and the $U(1)$ Ward identity of elementary quantum field theory.  The Chern-Simons equation of motion is $F_{\m\n}=0$, so the bulk-to-boundary propagator is a total
derivative  with respect to the bulk coordinates. The general structure is $K_{\mu,i}(x,\vec{w})= \pa_\m \L_i(x,\vec{w})$ where $x$ and $\vec w$ denote bulk and boundary points respectively. The specific form found in \cite{Dong:2014tsa} is\footnote{Throughout this paper we use the Poincar\'e patch metric $ds^2 =(L^2/x_0^2)(dx_0^2 + d\vec x\cdot d\vec x).$}

\begin{eqnarray}
  \label{eq:BuBoCS}
K_{\mu i}(x,\vec z)=\frac{\partial}{\partial x^\mu} \Lambda_i(x,\vec
z)=\frac{\partial}{\partial x^\mu} \frac{2\epsilon_{0ij}(\vec x-\vec
z)^j}{x_0^2+(\vec x-\vec z)^2} \,.
\end{eqnarray}
This form must be modified by multiplication on the boundary index by the projection operators $(\d_{ij} \pm i\e_{ij})/2$ to enforce the (anti-)self-duality property of $A_\m,(\tilde A_\m).$

\begin{figure}[h]
\begin{center}
\includegraphics[width=0.2\textwidth]{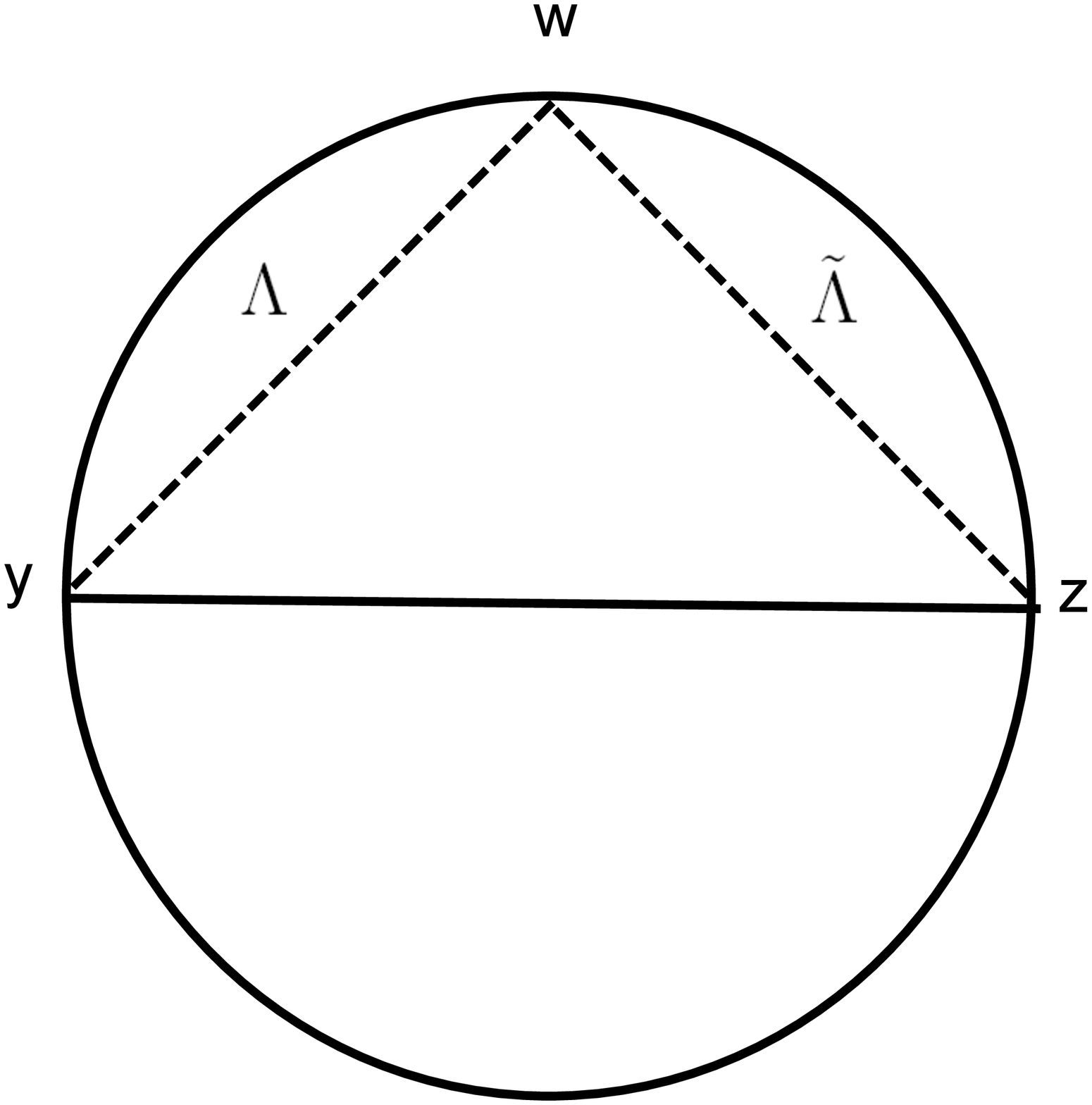}
\hspace*{0.35cm}
\includegraphics[width=0.2\textwidth]{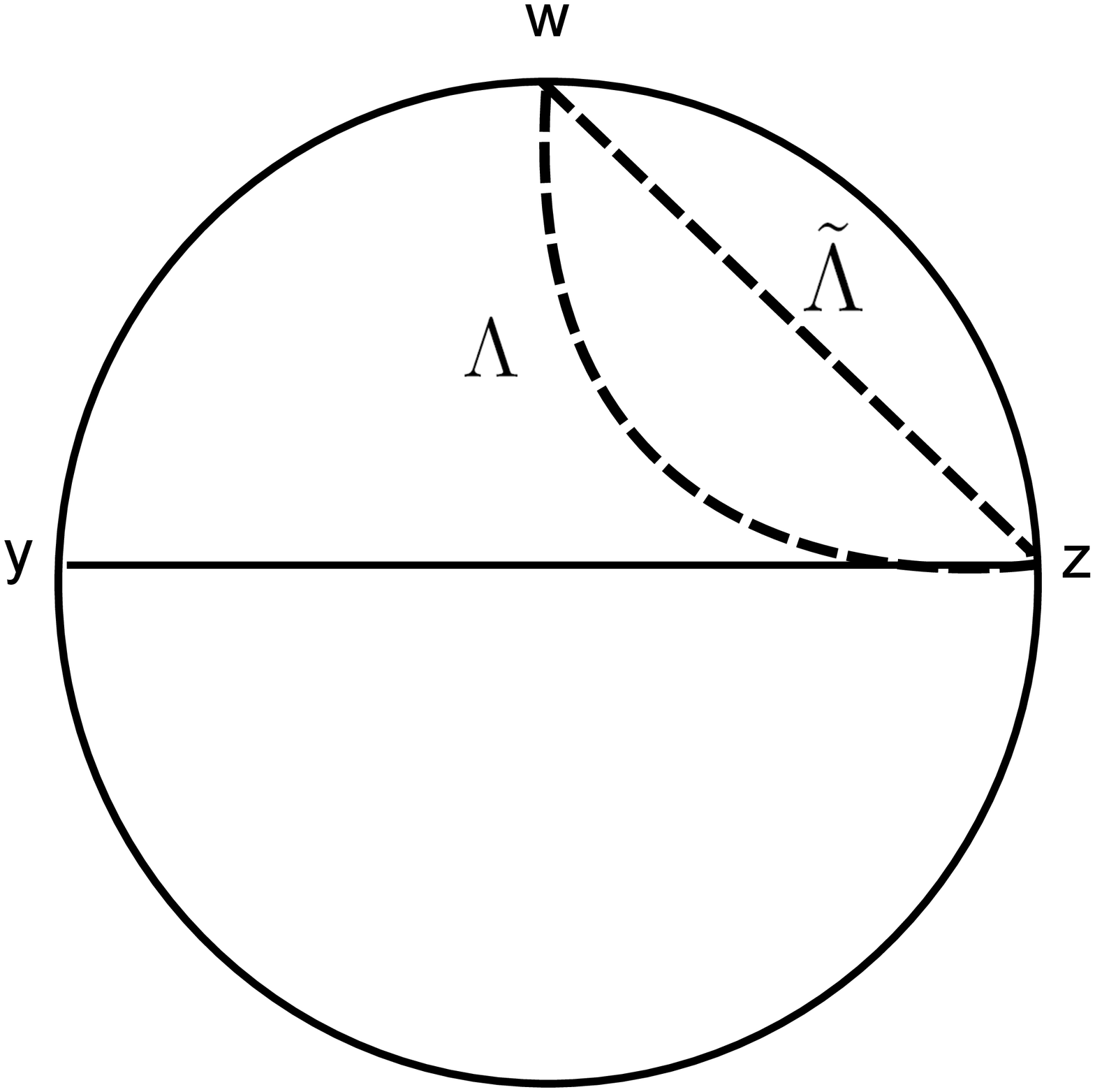}
\caption{Two of diagrams that depict the $\Lambda$-factors of the
photon pinned at the boundary after bulk integrals are performed.
There are also diagrams with $\vec y$ and $\vec z$ switched.}
\label{fig:bdy}
\end{center}
\end{figure}

Because the bulk-to-boundary propagator is a total derivative one
can integrate by parts in the $x$ and $x'$ integrals in Fig.
\ref{fig:Charged}. For each integral one finds a boundary term that
must be carefully studied plus a bulk term to which the Ward
identity (aka Green's theorem) applies.\footnote{A complete
calculation in a simpler example is given in Sec. \ref{sec:JOcOc}
below.  In the main, however, we refer readers to
\cite{Dong:2014tsa} for details of these straightforward but tedious
calculations.} A detailed calculation shows that only boundary terms
from partial integration contribute to the final answer; all bulk
terms cancel among the contributing diagrams. In the boundary terms
that remain, the $\L_i$ and $\tilde \L_i$ factors are pinned at the
end-points of the $\phi_c$ line as shown in Fig. \ref{fig:bdy}.

The $d^2w$ boundary integral then produces the final result
\begin{eqnarray}
  \label{eq:ChargeMassFinal}
\delta_h\langle O^\dag_c(\vec{y}) O_c(\vec{z})\rangle = ( 2\pi h q
\tilde{q}\ \textrm{log}\frac{|y-z|^2}{|a|^2} )\langle
O^\dag_c(\vec{y}) O_c(\vec{z})\rangle_0 \,,
\end{eqnarray}
where $a$ is the short-distance regulator. This indicates that the conformal dimension of $O_c$ changes by
\begin{equation}
\label{eq:Dimchange} \delta_h\Delta_{O_c} = -2\pi h q \tilde q \,.
\end{equation}
A similar result is  obtained for the change of scaling dimension of the spinor operator $\Psi_c$,  namely
\be \label{spinopdim}
 \delta_h\Delta_{\Psi_c} = -2\pi h (q-1) \tilde q \,.
\ee This example illustrates a basic feature of our results.  We
find SUSY breaking shifts of scaling dimension and bulk mass simply
because the R-charges of boson and fermion components of a
supermultiplet are different.

\subsubsection{R-neutral particle}\label{sec:R-neutralMass} In this
section, we outline the calculations which show that the mass of an R-neutral field such as the  modulus $\phi_m$ is not shifted by
the SUSY breaking term  (\ref{eq:SUSYBreak}).

Unlike R-charged particles, $\phi_m$ is  R-neutral and  thus does not couple \emph{directly}  to R-gauge bosons.
However,  $\phi_m$ does couple to
$\phi_c\phi_c^*$ pairs via the cubic coupling in Eq. (\ref{eq:cubic}). Thus  SUSY breaking effects might propagate to $\phi_m$ through
loops.  In the MSSM this effect is a major contributor to the shift in the Higgs boson mass.  In our theory the $U(1)$ Ward identity saves the day, and  $\phi_m$ remains massless to all perturbative orders.

\begin{figure}[h]
\begin{center}
\includegraphics[width=0.2\textwidth]{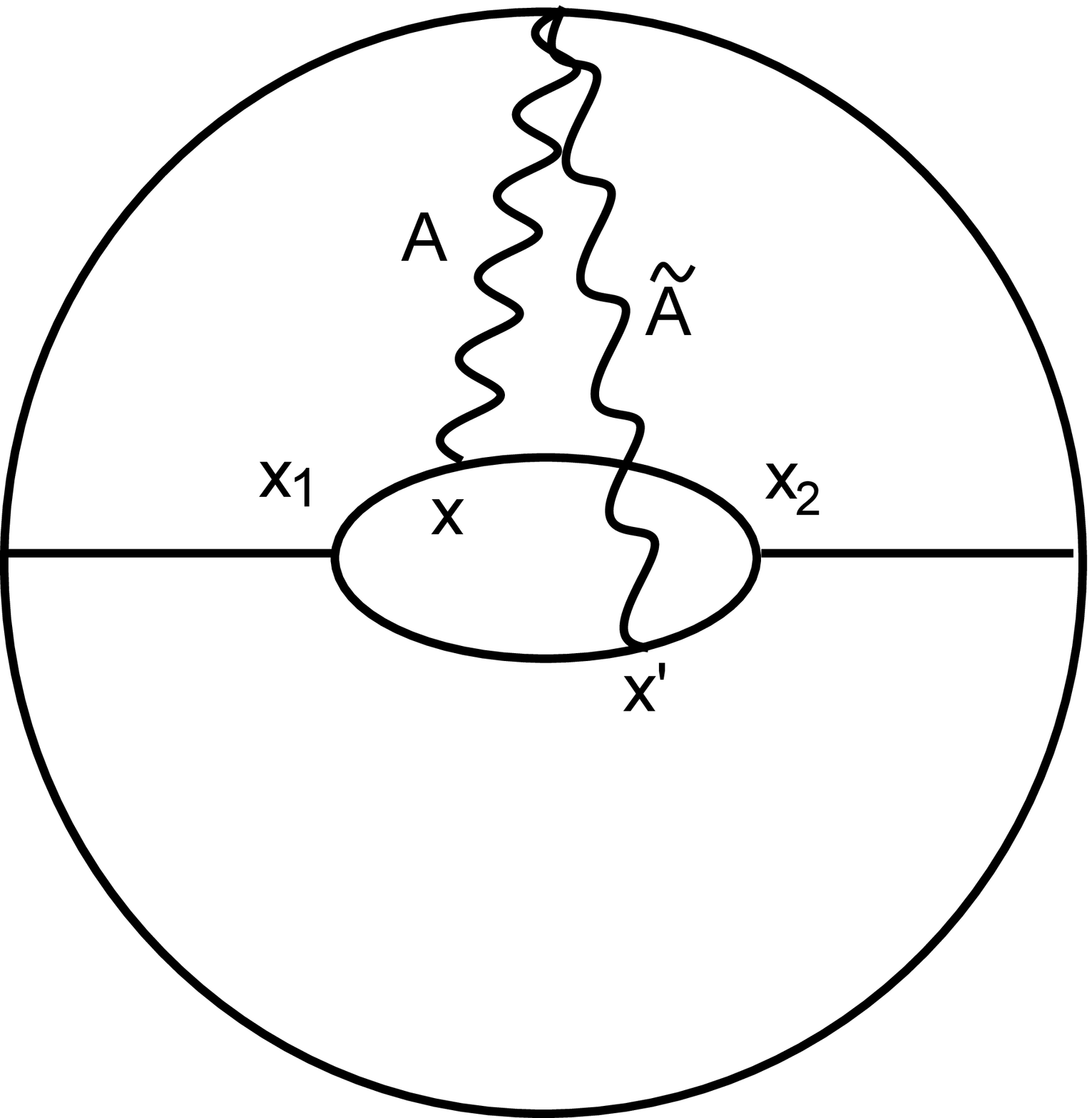}
\hspace*{1.35cm}
\includegraphics[width=0.2\textwidth]{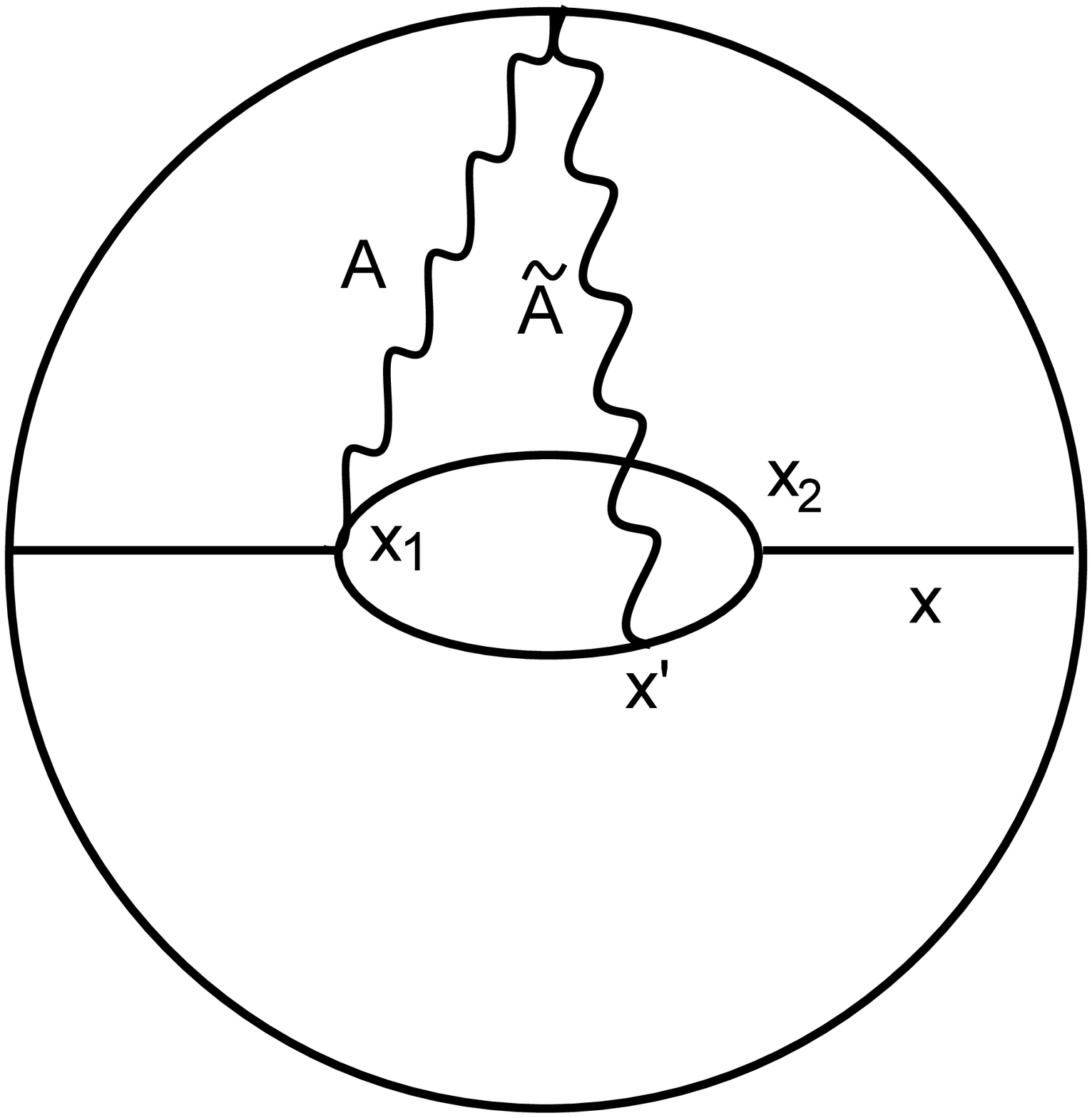}
\caption{Two of several diagrams that contribute to the leading
order deformation of $\<O_mO_m\>$.  The quartic vertex in the second diagram comes from the covariant derivative in (\ref{eq:cubic}).}
 \label{fig:LoopLeading}
\end{center}
\end{figure}

Two of several contributing one-loop Witten diagrams are shown in
Fig. \ref{fig:LoopLeading}. As in the previous section  one can use
 integration by parts and the Ward identity to evaluate  bulk integrals.  Since $\phi_m$ does not couple to  $A$ or $\tilde{A}$, the $\L_i$ factors cannot reach the  endpoints on the $AdS$ boundary. Instead they simply cancel among all diagrams. Indeed a careful calculation in \cite{Dong:2014tsa}  shows that the sum of all diagrams vanishes, so that
 $\phi_m$ remains massless.

The same cancellation can be  proven diagrammatically at any loop
order. Again it is  a consequence of the structure $K_{\mu i} = \pa_\mu \L_i$ and the Ward
identity.  Since there are no external $\phi_c$ lines,
the  boundary terms from  partial integration at each vertex vanish (provided that the unitarity bound $\D_c \ge 0$ for $AdS_3$ is satisfied). The $\L_i$ factor from any given bulk vertex is pinned at the adjacent vertices and these terms cancel among all diagrams. This is just the $U(1)$ charge conservation in  the action!

Analogous cancellations occur for n-$pt$ correlation functions of
$\phi_m$, which indicates the self-potential of $\phi_m$ remains
exactly flat.

\subsection{Coupling Constants} Supersymmetry not only requires
mass relations between superpartners but also specific relations among coupling
constants. In this section, we calculate the change of coupling
constants induced by the SUSY breaking term in Eq.
(\ref{eq:SUSYBreak}). We will show that the coupling constants are
shifted in a non-supersymmetric way. To simplify the calculation,
let us focus on the cubic coupling in Eq. (\ref{eq:cubic}). See Sec. 6 of
\cite{Dong:2014tsa} for more detail.

From $AdS/CFT$, the cubic coupling is related to the three point
correlation function $\langle O^\dag_c O_c O_m\rangle$. By conformal
invariance, it can written as
\begin{eqnarray}
  \label{eq:CFT3pt}
\langle O^\dag_c(\vec{y}) O_c(\vec{z}) O_m(\vec{w}))\rangle
=\frac{c_3}{|\vec{y}-\vec{z}|^{2\Delta_c-\Delta_m}
|\vec{y}-\vec{w}|^{\Delta_m} |\vec z-\vec w|^{\Delta_m}} \,.
\end{eqnarray}
When we turn on the SUSY breaking term in Eq. (\ref{eq:SUSYBreak}),
both $c_3$ and $\Delta_c$ are modified. After properly normalizing
$O_c$, the change of $c_3$ can be written as
\begin{eqnarray}
  \label{eq:c3Change}
\delta_h c_3 = -\frac{4\pi q \tilde{q} h}{\Delta_c-1} c_3 \,.
\end{eqnarray}
If the coupling   (\ref{eq:cubic}) is the only cubic term in the bulk action,  then the Witten diagram which
determines the 3-$pt$ correlation function is given by
\begin{eqnarray}
  \label{eq:3ptADSCFT}
\langle O^\dag_c(\vec{y}) O_c(\vec{z}) O_m(\vec{w}))\rangle =
-\lambda \int \frac{d^3x}{x_0^3}\pa_\m K_{\Delta_c}(x,\vec{y})
\pa^\m K_{\Delta_c}(x,\vec{z}) K_{\Delta_m}(x,\vec{w}) \,.
\end{eqnarray}
It is straightforward to check that the leading order correction to
$K_{\Delta_c}$ can be simply accounted by shifting $\Delta_c$ to
$(\Delta_c-2\pi q \tilde{q} h)$, i.e.
\begin{eqnarray}
  \label{eq:KLeading}
K_{\Delta_c,h}(x,\vec{x}')=K_{\Delta_c-2\pi q \tilde{q}
h}(x,\vec{x}') \,.
\end{eqnarray}
After a careful calculation, we find
\begin{eqnarray}
  \label{eq:lambdaChange}
\frac{\delta\lambda}{\lambda}=2\pi q \tilde{q}h
\frac{\partial}{\partial \Delta_c}
\textrm{log}\frac{\Gamma(\Delta_c-\frac{\Delta_m}{2})\Gamma(\Delta_c+\frac{\Delta_m}{2}-1)}{\Gamma(\Delta_c)^2}(\D_c^2 -2\D_c -\frac12\D_m^2+\D_m)\,.
\end{eqnarray}
Thus we see that the shifts of coupling constants depend on R-charges.
This is particularly interesting because it indicates
that supersymmetric relations between coupling constants are also broken
when the SUSY breaking deformation is turned on.

However, there is one subtlety that we need to address at the end of
this section. In the calculation above we make the assumption that
the cubic vertex in Eq. (\ref{eq:cubic}) is the only interaction
vertex contributing to the 3-$pt$ correlation function after SUSY is
broken. However, it is  possible that higher derivative bulk
couplings, such as $\lambda' (\nabla^\mu \nabla^\nu
\phi_c^\dag)(\nabla^\mu \nabla^\nu \phi_c) \phi_m$, can be induced
when $h$ is turned on. These vertices will also contribute to the
3-$pt$ correlation function. We expect a careful calculation of
4-$pt$ functions may resolve this ambiguity. However that is beyond
the scope of this paper.
 It is sufficient for our purpose here to show that the
SUSY relation among coupling constants is explicitly violated.

\subsection{Singly charged particles}\label{sec:SingleCharge}
In this section, we consider SUSY breaking effects on particles
that are charged  under only one of the R-symmetry groups.  We choose bulk fields with charge $(q,0)$. This simplification is especially helpful since it
provides important intuition for the extension of our
mechanism to higher dimensions.

The order $h$ correction discussed in Sec. 2.1.1 vanishes  for a singly charged
particle, but there is an order $h^2$  shift of conformal dimension which was calculated by CFT methods in \cite{Dong:2014tsa} with the result
\begin{equation}
\label{eq:Dimchange2} \delta_{h^2}\Delta = \pi^2k h^2 q^2/2 \,.
\end{equation}
Thus SUSY breaking effects also occur for particles with
only one non-vanishing R-charge.

We now assume that \emph{all} charged fields in the theory couple
only to $A_\m$. The gauge field  $\tilde A_\m$ then appears only in
the Chern-Simons term (\ref{eq:CSterm}) and in the SUSY breaking
term (\ref{eq:SUSYBreak}). One can then integrate out $\tilde{A}$
and obtain an  action with the new non-local  SUSY breaking boundary
term
\begin{eqnarray}
  \label{eq:SUSYBreakSingle}
S'_{SB}=-\frac{h^2 \tilde k}{8}\int_{bdy}d^2 w_1 d^2 w_2 A_z(w_1)
\frac{J^{zz}(w_1,w_2)}{|w_1-w_2|^2}A_z(w_2) \,.
\end{eqnarray}
One can now apply the basic partial integration plus Ward identity methods to the Witten diagrams that describe the correction of this non-local interaction to the correlator $\<O_cO_c^*\>.$ This calculation reproduces the result (\ref{eq:Dimchange2}) exactly.

Integrating out a light field always produces a non-local effective
action, so there is no fundamental problem with $S'_{SB}$.  However
it is worthwhile to discuss how the common concern that a non-local
operator violates causality is avoided  operationally in our
framework. As we have seen in earlier sections, the net result of
the calculations is that the $\L_i$ factors of the gauge field
propagators are pinned at the boundary points of R-charged fields.
The same feature occurs for insertions of $S'_{SB}$. This means that
the original correlation function appears as a factor in the final
amplitude. Thus if two points are causally disconnected in the
original theory, i.e. the commutator at these two
points vanishes, these two points are still causally disconnected in
the deformed theory.

\subsection{All-order sum of SUSY breaking corrections}\label{sec:SumAdS3}
\begin{figure}
\begin{center}
\includegraphics[width=0.2\textwidth]{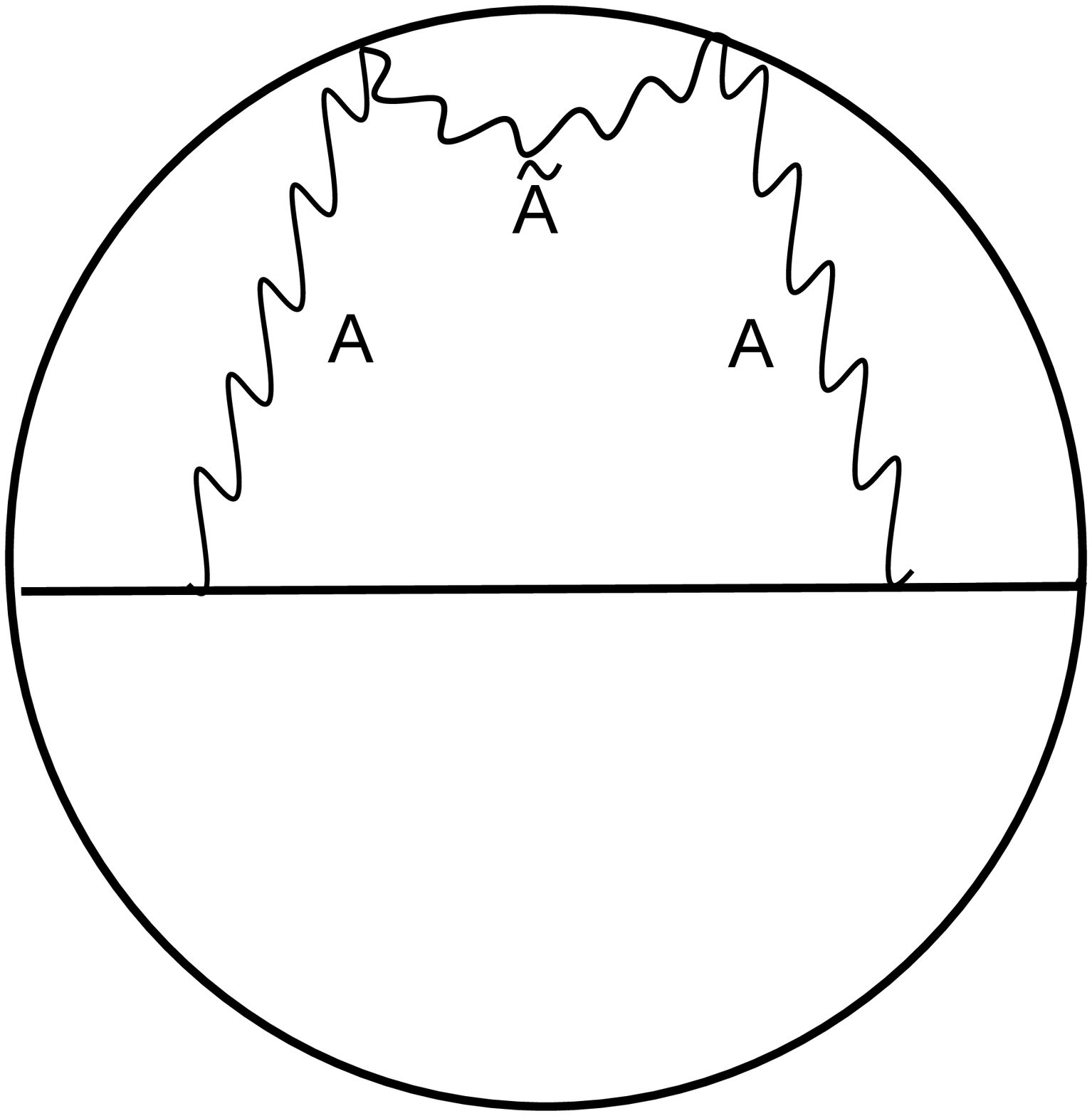}
\hspace*{1.35cm}
\includegraphics[width=0.2\textwidth]{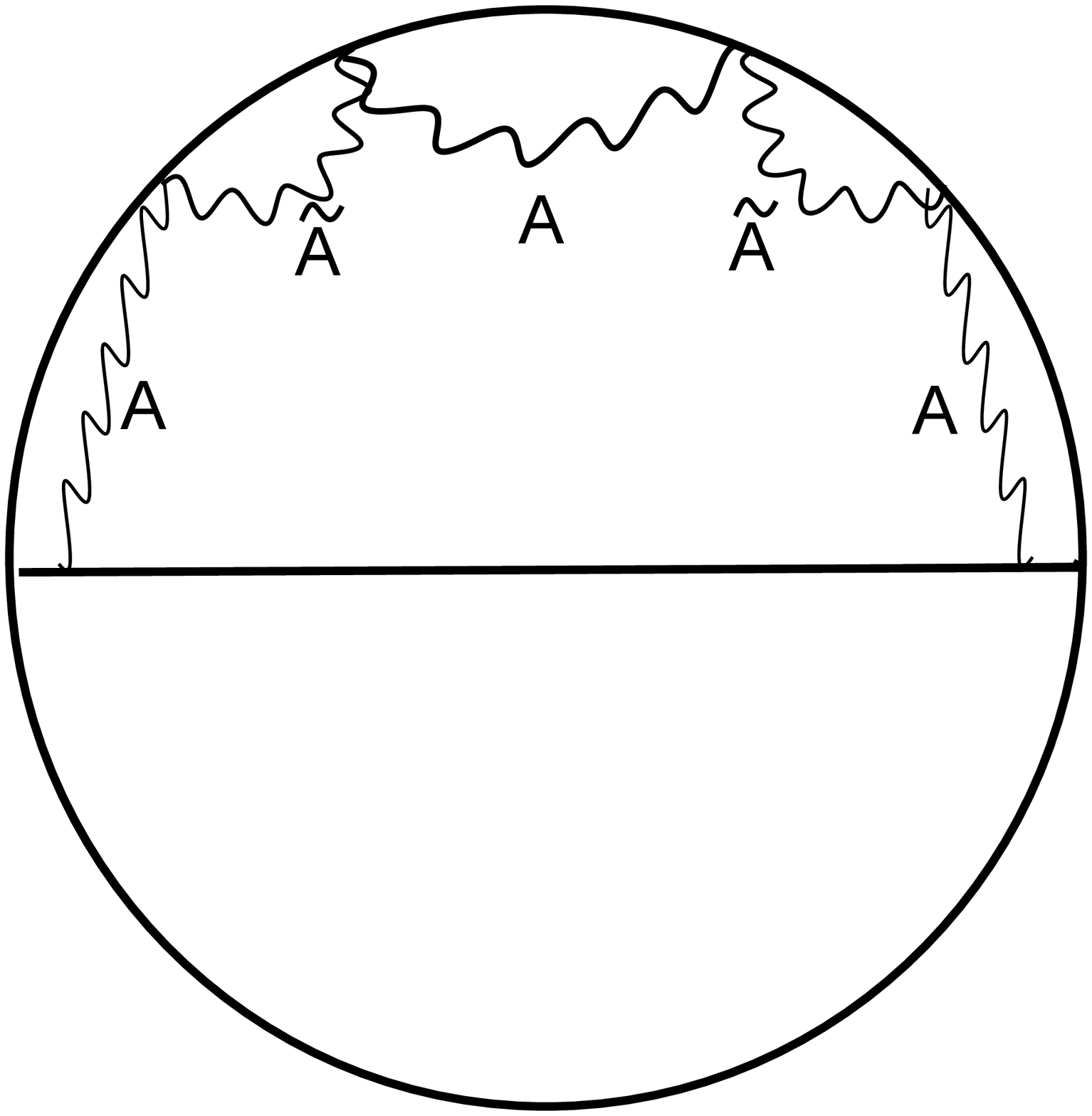}
\caption{The diagram on the left describes the effective order $h^2$ SUSY breaking discussed in Sec.\ref{sec:SingleCharge}.
The diagram on the right contributes the $h^4$ term in the expansion of (\ref{eq:AdS3Precise}). }
 \label{fig:LoopAll}
 \end{center}
\end{figure}

Since SUSY is broken by a bilinear operator, one might expect to obtain an exact solution for SUSY breaking corrections which supersedes the order $h$ and $h^2$ contributions discussed so far.  We  obtained the exact solution by summing the diagrams shown in  Fig. \ref{fig:LoopAll}.    The result for the shifted scaling dimension of an operator with general R-charges $(q,\tilde q)$ is presented in \cite{Dong:2014tsa}. For the  slightly simpler case of charge $(q,0)$, the shifted scaling dimension is
\begin{eqnarray}
  \label{eq:AdS3Precise}
\Delta =\Delta_0 + \frac12\,\frac{ \pi^2h^2k q^2}
{\,1-\pi^2h^2 k^2/4} \,.
\end{eqnarray}
As the coupling $h$ is increased toward the root of the denominator
the difference in conformal dimensions between $\phi_c$ (with $q^2$
in the numerator) and $\psi_c$ (with $(q-1)^2$) becomes arbitrarily
large.

\begin{figure}
\begin{center}
\includegraphics[width=0.2\textwidth]{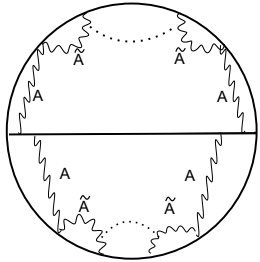}
\caption{This diagram contributes a shift of the power law in the
corrected to 2-pt correlation function.}
 \label{fig:LoopAll2}
 \end{center}
\end{figure}

The diagrams  of Fig. \ref{fig:LoopAll} were summed by deriving a
recursion relation which reduces higher $h$ order corrections to the
lowest order. This recursion relation requires the identity
\begin{eqnarray}
  \label{eq:recursionAdS3}
\partial_i\langle J^i(\vec x) J^j(\vec y)\rangle = -\pi \frac{\pa}{\pa y_j}
\delta^{(2)}(\vec x-\vec y) \,.
\end{eqnarray}
This identity is special to
$d=2$. In higher
dimensions, $\partial_i\langle J^i(\vec x) J^j(\vec y)\rangle$ is
precisely zero, and we will obtain a very different all-order formula in later sections.

The two diagrams in Fig. \ref{fig:LoopAll} are the lowest terms of a series of diagrams of order $h^{2n}q^2$.  The summed series may be thought of as a "necklace" that determines the all-order shift of the conformal dimension (\ref{eq:AdS3Precise}). Fig. \ref{fig:LoopAll2}  depicts the insertion of two independent necklaces along the charged line.  As discussed in Sec. 5 of \cite{Dong:2014tsa},  the partial integration and Ward identity arguments apply to each necklace independently. This produces an exponential series so that the fully corrected correlator $\<O_c^\dagger O_c\>$ acquires the shifted power law required by conformal symmetry:
\be \label{ococexactd2}
\langle O_c^{\dagger}(\vec y) O_c(\vec z)\rangle
\sim \frac{1}{(\vec y - \vec z)^{2\D}}\,.
\ee

\section{The undeformed correlator $\<J^i(\vec z) O_c^\dagger(\vy) O_c(\vec w)\>$}\label{sec:JOcOc}
 \setcounter{equation}{0}
This is a transitional section from $AdS_3$ to  $AdS_{d+1}$.  We will study the 3-point function $\<J^i(\vec z) O_c^\dagger(\vy) O_c(\vec w)\>$ in the undeformed theory.
This will illustrate the basic partial integration and Ward identity technique in a  simple example that will provide a very  important clue to the higher dimensional extension of the $AdS_3$ model.

The previous discussion has made clear the importance of the total derivative structure $K_{\mu i}(\vec w)= \pa_\mu \L_i(x,\vec w)$. This property follows immediately from  Chern-Simons dynamics in which $F_{\m\n} = 0 $ is the classical equation of motion. This suggests that one extension to higher dimensions might be based on a
 $BF$ theory with Lagrangian
 \begin{eqnarray}
  \label{eq:BF}
L\sim  B\wedge F \,.
\end{eqnarray}
For  $AdS_{d+1}$, the field $B$ is a $(d-1)$-form and the field strength $F=dA$ with $A$  a 1-form.  However this suggestion is not viable because the 3-point function to which it leads agrees with the unique conformal invariant form
(see (23.54) of \cite{bible})
\begin{eqnarray}
  \label{eq:JOOCFT}
\langle J_i(\vec z) O_c^\dag (\vec y) O_c(\vec w)\rangle\sim
\frac{1}{(\vec y - \vec w)^{2\Delta_c-d+2}(\vec y -\vec
z)^{d-2}(\vec w-\vec z)^{d-2}}[\frac{(\vec y-\vec z)_i}{(\vec y-\vec
z)^2}-\frac{(\vec w-\vec z)_i}{(\vec w-\vec z)^2}]
\end{eqnarray}
only for $d=2$.

\begin{figure}
\begin{center}
\includegraphics[width=0.2\textwidth]{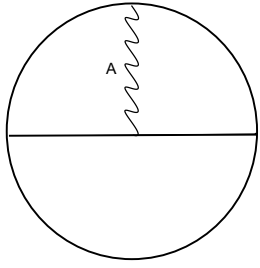}
\caption{Witten diagram for  $\langle J^i O_c^\dag
O_c\rangle$}
 \label{fig:JOO}
 \end{center}
\end{figure}

To demonstrate this let us evaluate the Witten diagram of Fig. \ref{fig:JOO} for general boundary dimension $d$.  We assume that the bulk-boundary propagator of the gauge field dual to the conserved current $J^i(\vec w)$ has  the total derivative form $K_{\mu i}(\vec w)= \pa_\mu \L_i(x,\vec w)$,
with $\L_i$ to be specified later.  We also need the bulk-boundary propagator for the bulk scalar $\phi_c(\vec x)$,  namely
\bea\label{blkbdyphic}
K_\D(x_0,\vec x) &=& c_\D\bigg(\frac{x_0}{x_0^2 +\vec x^2}\bigg)^\D \,,\qquad\quad   (\Box- m^2)K_\D(x_0,\vec x)=0 \,,\\
\lim_{x_0\to 0} K_\D(x_0,\vec x) &=& x_0^{d-\D}\d^{(d)}(\vec x) \,, \qquad\quad C_\D =\frac{\G(\D)}{\pi^{d/2}\G(\D-d/2)} \,.
\eea

The value of the Witten diagram is given by the bulk integral
\bea
\langle J^i(\vec w) O_c^\dag(\vec y)O_c(\vec z)\rangle &=& \int d^{d+1}x\sqrt{g}\pa_\m\L_i(x,\vec w)K_\D(x_0,\vec x-\vec y) \overset{\leftrightarrow}{\pa^\m} K_\D(x_0,\vec x-\vec z)\\
&=&- \int d^{d+1}x\sqrt{g}\L_i(x,\vec w)K_\D(x_0,\vec x-\vec y) \overset{\leftrightarrow}{\Box}  K_\D(x_0,\vec x-\vec z)\\
&-& \lim_{x_0\to 0}\int d^dx \sqrt{g} K_\D(x_0,\vec x-\vec y)
\overset{\leftrightarrow}{\pa^0} K_\D(x_0,\vec x-\vec z)\,. \eea
After partial integration and use of Green's theorem we find the
bulk integral in the second line (which vanishes by the equation of
motion for $K_\D$) plus the boundary term in the third line.  The
limit $x_0\to 0$ of the boundary term vanishes due to the $x_0$
factors in $K_\D(x_0,\vec x)$ except when $\vec x \approx \vec y$
and $\vec x \approx \vec z$.  The contribution from these regions is
determined by the limit of $K_\D$ given in (\ref{blkbdyphic}) above.
We find
\begin{eqnarray}
  \label{eq:JOObulk}
\langle J^i(\vec w) O_c^\dag (\vec y) O_c(\vec z)\rangle = (\D-d)
[\Lambda_i(\vec z-\vec w)-\Lambda_i(\vec y-\vec w)]\frac{1}{(\vec y
- \vec z)^{2\Delta_c}} \,.
\end{eqnarray}
We can now compare this result with the required form
(\ref{eq:JOOCFT}). In our $AdS_3$ example, the R-gauge boson
propagator is a total derivative as Eq. (\ref{eq:BuBoCS}), and we
find agreement with (\ref{eq:JOOCFT}) when its $\L_i$ factor is
inserted in (\ref{eq:JOObulk}). For $d\ge3$, however, no choice of
$\L_i(\vec z-\vec w)$  can bring (\ref{eq:JOObulk}) and
(\ref{eq:JOOCFT}) into agreement because it fails to produce the
correct parametric coordinate dependence .

Thus we pass to the next section with an acute puzzle. We cannot use
a gauge field of the form $K_{\mu i} = \pa_\m \L_i$ in the undeformed
current-matter correlator, yet we need precisely such a form to use
Ward identity methods to evaluate SUSY breaking corrections. Please
read on to see how this dilemma is resolved.

\section{Generalization to $AdS_{d+1}$}\lab{AdS4}
 \setcounter{equation}{0}
The puzzle in Sec. \ref{sec:JOcOc} motivates us to proceed as follows towards the desired higher dimensional model:

a. We use a gauge field $A_\mu$ with Maxwell dynamics which is dual to a conserved $U(1)_R$ current $J_i$ with correct undeformed correlation functions. A single gauge field suffices.

b. Its bulk-boundary propagator $K_{\mu i}$ is not a total $\pa_\m$
derivative, but we find a simple identity which shows that $\pa_i
K_{\m}{}^i$ is a total $\pa_\m$ derivative.

c. We employ a SUSY breaking boundary term which always produces $\pa_i K_{\m}{}^i$ when inserted in a Witten amplitude.  Thus the same powerful
Ward identity methods used in the $AdS_3$ model are also valid in higher dimensions.

d. The SUSY breaking term is non-local, but its structure is similar to that of Sec. \ref{sec:SingleCharge} and has conventional causal properties. It is also exactly marginal.

\subsection{Maxwell theory in $AdS_{d+1}$ space}\label{sec:MaxAdS}

The well-known Maxwell  action  is
\begin{eqnarray}
  \label{eq:MaxAction}
S_{Maxwell}=\int d^{d+1}x \sqrt{g}
\frac{1}{4}F_{\mu\nu}F^{\mu\nu} \,.
\end{eqnarray}
A solution of the classical equation of motion in $AdS_{d+1}$ has the  near-boundary expansion
\begin{eqnarray}
  \label{eq:AsympExp}
A_i(x)=\alpha_i(\vec x)(1+ \ldots) +\beta_i(\vec x) (x_0^{d-2}+...) \,.
\end{eqnarray}
We use standard quantization in which $\a_i(\vec x)$ is fixed as the boundary value of the bulk field and $\b_i(\vec x)$ is the response, which can be thought of as the current  operator $J_i(\vec x)$ in the dual $CFT_d$. Note that
$\b_i(\vec x)$ has scaling dimension $d-1$ and that $\pa_i\b^i =0$ as a consequence of the Maxwell equation.
The bulk-to-boundary propagator of the Maxwell field in $AdS_{d+1}$
satisfies the sourceless Maxwell equation. It is gauge dependent,
and we use the conformal covariant form (see (48) of
\cite{Freedman:1998tz})
\begin{eqnarray}
  \label{eq:BuBoMax}
K_{\mu i}= \frac{\Gamma(d)}{2\pi^{d/2}\Gamma(d/2)} \frac{J_{\mu
i}(x-\vec z) x_0^{d-2}}{[x_0^2+(\vec x-\vec z)^2]^{d-1}}\,.
\end{eqnarray}
Standard quantization of Maxwell theory is well defined for
$d>2$. It is important to realize that this bulk-to-boundary
propagator cannot be written as a total derivative. This is as
expected because $K_{\mu i}$ satisfies the Maxwell equation $\nabla^\m F_{\m\n} =0$, which does
not imply $F_{\m\n}=0$.

\subsection{The identity} \label{sec:identity}
In order to convert this bulk-to-boundary propagator to an object
useful for our purposes, we need an identity.  We first present it
as a general property of bulk-to-boundary propagators for $n$-form gauge
fields with Maxwell dynamics in $AdS_{d+1}$, namely
\begin{eqnarray}
  \label{eq:Identity}
K_{\mu_1...\mu_n}^{\ \ \ \ \ \ \ i_1...i_n}(x,\vec z)&\equiv&
\frac{\Gamma(d)}{2\pi^{d/2}\Gamma(d/2)}
(x_0)^{d-2n}\frac{J_{\mu_1}^{\ \ [i_1}...J_{\mu_n}^{ \ \
i_n]}}{(x_0^2+(\vec x-\vec z)^2)^{d-n}} \,,\nonumber\\
\partial_{i_1}K_{\mu_1...\mu_n}^{\ \ \ \ \ \ \
i_1...i_n}&=&\frac{2d-2n+2}{d-2n+2}\partial_{[\mu_1}K_{\mu_2...\mu_n]}^{\
\ \ \ \ \ \ \ i_2...i_n} \,.
\end{eqnarray}
In this paper we need only the case $n=1$. This gives us  a very useful identity, because it converts the bulk-to boundary propagator in Maxwell theory
to a total derivative respect to bulk coordinates, i.e.
\begin{eqnarray}
  \label{eq:IdentityMax}
\partial_{i}K_{\mu}^{\ \ i}(x,\vec
z)=\frac{\Gamma(d)}{\pi^{d/2}\Gamma(d/2)}\partial_{\mu}\frac{x_0^d}{(x_0^2+(\vec
x-\vec z)^2)^d} =\pa_\m  K(x,\vec z) \,,
\end{eqnarray}
where $K$ is  the bulk-boundary propagator of a massless scalar.
This identity will be a  critical tool in the evaluation of SUSY breaking corrections in higher dimension $d \ge 3$.

Here is a qualitative interpretation of the identity.  The boundary
integral of $K_{\mu i}$ computes the bulk field $A_\mu(x)$ with boundary value
$\a_i$. In general there is a non-vanishing $F_{\m\n}. $
The $\pa_i$ in the identity means that $\a_i$  is purely
longitudinal. In this case  one expects that $F_{\m\n}$ vanishes and
that $A_\m$ is a total derivative.

\subsection{SUSY breaking term in $AdS_{d+1}$}\label{sec:SUSYAdS}
The generalization of our SUSY breaking
mechanism to higher dimensional $AdS$ space is motivated by the second-order SUSY
breaking action in $AdS_3$, given in  (\ref{eq:SUSYBreakSingle}).  We
propose that the general SUSY breaking action in the CFT language is
\begin{eqnarray}
  \label{eq:SUSYBreakCFTHigh}
S_{CFT_{d}} &=& h\int d^d \vec w_1 d^d \vec w_2 J_i(\vec
w_1)\frac{J_{ij}(\vec w_1-\vec w_2)}{(\vec w_1-\vec w_2)^2}J_j(\vec
w_2) \,,\\
J_{ij}(\vec w) &\equiv& \d_{ij} - \frac{2w_iw_j}{\vec w^2}\,.
\end{eqnarray}
Using the BDHM dictionary \cite{Banks:1998dd}, we may write the CFT current $J_i(\vec w)$ as the limit of the bulk operator $cw_0^{2-d} A_i(w)$ as $w_0$ goes to zero, where $c$ is a constant.  The above deformation can therefore be written in the bulk language as
\begin{eqnarray}
  \label{eq:SUSYBreakHigh}
S_{AdS_{d+1}}=\lim_{w_{10},w_{20}=\ \epsilon\rightarrow 0} c^2 h\int
d^d \vec w_1 d^d \vec w_2 (w_{10}^{2-d})(w_{20}^{2-d}) A_i(w_1)\frac{J_{ij}(\vec w_1-\vec w_2)}{(\vec w_1-\vec w_2)^2}A_j(w_2) \,.
\end{eqnarray}
This action is written on the constant radial coordinate slice $w_{10}=w_{20}= \epsilon$, and the boundary limit is taken after its insertion in a correlation function.  For future use, $\frac{J_{ij}(\vec w)}{\vec
w^2}$ can be written in the more convenient form
\begin{eqnarray}
  \label{eq:DoubleLog}
\frac{J_{ij}(\vec w)}{\vec w^2} =
\frac{1}{2}\partial_{i}\partial_{j} \textrm{Log}(\vec w^2)\,.
\end{eqnarray}
In later calculations,  partial integration of  $\pa_i$ will convert
the bulk-to-boundary propagator $K_{\mu i}$ to a total derivative.
Naively, if one substitutes Eq. (\ref{eq:DoubleLog}) to Eq.
(\ref{eq:SUSYBreakCFTHigh}) and integrates by parts, it seems that
the SUSY breaking term vanishes due to current conservation. We emphasize again that unlike ordinary QFT in flat spacetime,
one should not use current conservation or the equation of motion to remove a term
in the CFT Lagrangian or in a theory living in AdS space. This subtlety
already appears in the SUSY breaking term that we introduced in our $AdS_3$
model, as discussed in Sec. \ref{AdS3}. In the following discussion, we
will present the detailed calculation to show that this SUSY
breaking term generates non-trivial physical effects, and its
non-triviality is consistent with the Ward identity from the CFT point of
view.

It is worthwhile to understand the proposal (\ref{eq:SUSYBreakHigh}) better  before applying it to calculate  SUSY breaking effects.  First we note that
$S_{AdS_{d+1}}$ is invariant under dilatations $w_{i\mu} \to  \l w_{i\mu}$ and $A_i(\vec w_i)\to \l^{-1}A_i(\l \vec w_i)$.  This means that it acts as a marginal deformation of the boundary $CFT_d$.  Indeed we will verify from its effect on correlation functions that it is exactly marginal.

We will compare  the  SUSY breaking deformation of a correlation
function from both the insertion of $S_{AdS_{d+1}}$ and of
$S_{CFT_{d}}$.  In the first case we compute the bulk and boundary
integrals using Ward identity methods. In the second case we use
conformal perturbation theory. The results agree.

The SUSY breaking actions are non-local.  They have the same
structure as the term  (\ref{eq:SUSYBreakSingle}) used in $AdS_3$.
However, as far as we know, there is no way to rewrite them as the
result of integrating out massless degrees of freedom in a larger
theory. Nevertheless, as we discussed in Sec. \ref{sec:SingleCharge},
we can check by explicit computation of SUSY breaking deformations
of correlation functions that no pathology occurs, i.e. the
casuality properties are exactly the same as undeformed theory.

\subsection{SUSY breaking corrections}\label{sec:SUSYAdSCorrection}
In this section, we examine physical effects of the SUSY breaking
term of Sec. \ref{sec:SUSYAdS}. First we show that the 2-pt
correlation functions of R-charged particles receive non-trivial
corrections. Thus there are explicit  SUSY breaking effects  in the
particle mass spectrum. Next we will show that the masses of
R-neutral particles do not receive any SUSY breaking corrections.
Then we return to charged particles and show that SUSY breaking
effects vanish beyond leading order $h$.  Finally we consider Witten
diagrams containing bulk gravitons and show that SUSY breaking
corrections vanish.

\subsubsection{Mass corrections for R-charged and R-neutral particles}\label{sec:SUSYAdSCorrectionRcharged}

\begin{figure}[h]
\begin{center}
\includegraphics[width=0.2\textwidth]{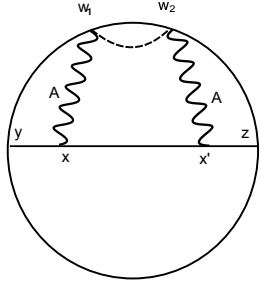}
\caption{A Witten diagram that gives an  order $h$ SUSY breaking
correction to the 2-pt correlation function of R-charged particle.}
 \label{fig:ChargedLeadingHigh}
\end{center}
\end{figure}

In this section, we calculate the leading order SUSY breaking correction
to the 2-pt correlation function of R-charged particle. The Witten
diagram to be computed is shown in Fig.
\ref{fig:ChargedLeadingHigh}. The effect of another diagram with the $A_\mu A^\mu\phi_c\phi_c^\dagger$ seagull vertex is included in the final result.

The diagram can be calculated as follows:
\begin{eqnarray}
  \label{eq:ChargedLeadingHigh}
\delta_h\langle O^\dag_c(\vec{y}) O_c(\vec{z})\rangle &=& -\frac{h
q^2}{2} \int d^d w_1 d^d w_2\ d^{d+1}x\
d^{d+1}x'\sqrt{g(x)}\sqrt{g(x')}
K_{\mu i}(x,\vec{w}_1)\frac{J_{ij}(\vec w_1-\vec w_2)}{(\vec w_1-\vec w_2)^2}K_{\nu j}(x',\vec{w}_2) \times\nonumber\\
&&\ \ \ \ \  \times[K_{\Delta}(x,\vec{y})\overset{\leftrightarrow}{\partial^{\mu}}(G_\Delta(x,x')\overset{\leftrightarrow}{\partial'^{\nu}}K_{\Delta}(x',\vec{z}))]+ (\vec{y}\leftrightarrow\vec{z})\nonumber\\
&=& \frac{h q^2}{4} \int d^d w_1 d^d w_2\ d^{d+1}x\
d^{d+1}x'\sqrt{g(x)}\sqrt{g(x')}
\partial_i K_{\mu i}(x,\vec{w}_1)\textrm{Log}[(w_1-w_2)^2]\partial_j K_{\nu j}(x',\vec{w}_2) \times\nonumber\\
&&\ \ \ \ \  \times[K_{\Delta}(x,\vec{y})\overset{\leftrightarrow}{\partial^{\mu}}(G_\Delta(x,x')\overset{\leftrightarrow}{\partial'^{\nu}}K_{\Delta}(x',\vec{z}))]+ (\vec{y}\leftrightarrow\vec{z})\nonumber\\
&=& \frac{h q^2}{4} (\frac{\Gamma(d)}{\pi^{d/2}\Gamma(d/2)})^2\int
d^d w_1 d^d w_2\ d^{d+1}x\
d^{d+1}x'\sqrt{g(x)}\sqrt{g(x')}\times\nonumber\\
&&\ \ \ \ \ \times\partial_{\mu}\frac{x_0^d}{(x_0^2+(\vec x-\vec
w_1)^2)^d}\textrm{Log}[(w_1-w_2)^2]\partial_{\nu}\frac{x_0^{'d}}{(x_0^{'2}+(\vec
x'-\vec w_2)^2)^d}\times\nonumber\\
&&\ \ \ \ \  \times[K_{\Delta}(x,\vec{y})\overset{\leftrightarrow}{\partial^{\mu}}(G_\Delta(x,x')\overset{\leftrightarrow}{\partial'^{\nu}}K_{\Delta}(x',\vec{z}))]+ (\vec{y}\leftrightarrow\vec{z}) \,.
\end{eqnarray}
Here in the second step, we use Eq. (\ref{eq:DoubleLog}) and
integrate by parts to move $\partial_i$ to $K_{\mu,i}(x,\vec w_1)$
and $\partial_j$ to $K_{\nu,i}(x',\vec w_2)$. The extra minus sign
is because $\partial_j$ now acts on $\vec w_2$ instead of $\vec
w_1$. In the last step, we use Eq. (\ref{eq:IdentityMax}) to convert
$\partial_i K_{\mu i}$ to a total derivative respect to $x_\mu$,
similar for $\partial_j K_{\nu j}$. One can integrate by parts on
$x$ and $x'$ integrals to simplify the calculation. The remaining calculation
is very similar to the one carried out for $AdS_3$, and the
result is the following:
\begin{eqnarray}
  \label{eq:ChargedLeadingHighFinal}
\delta_h\langle O^\dag_c(\vec{y}) O_c(\vec{z})\rangle &=& \frac{h
q^2}{4} (\frac{\Gamma(d)}{\pi^{d/2}\Gamma(d/2)})^2 \lim_{x_0,
x'_0\rightarrow a}\int d^d w_1 d^d w_2
d^{d}x'\nonumber\\
&&\ \ \ \ \ \frac{x_0^d}{(x_0^2+(\vec y-\vec w_1)^2)^d}
\frac{x_0^{'d}}{(x_0^{'2}+(\vec
x'-\vec w_2)^2)^d}\ \textrm{Log}[(w_1-w_2)^2]\times\nonumber\\
&&\ \ \ \ \
\times 2\Delta\frac{C_\Delta^2}{C_{\Delta+1}}\left[\frac{\delta^{d}(\vec{x}',\vec{y})}{|\vec{x}'-\vec{z}|^{2\Delta}}-\frac{\delta^{d}(\vec{x}',\vec{z})}{|\vec{x}'-\vec{y}|^{2\Delta}}\right]+ (\vec{y}\leftrightarrow\vec{z})\nonumber\\
&=&
-\frac{\Gamma(\Delta)}{\Gamma(\Delta-d/2)}\frac{(\Delta-d/2)}{\pi^{d/2}}h
q^2\frac{1}{|\vec{y}-\vec{z}|^{2\Delta}}\textrm{Log}[\frac{(\vec
y-\vec z)^2}{a^2}]\nonumber\\
&=& -h q^2\textrm{Log}[\frac{(\vec y-\vec z)^2}{a^2}] \langle
O^\dag_c(\vec{y}) O_c(\vec{z})\rangle_0 \,.
\end{eqnarray}
Here $a$ is the UV cut-off. In the second step, we used
\begin{eqnarray}
  \label{eq:IdentityDelta}
\lim_{x_0\to 0}\left(\frac{x_0}{x_0^2+(\vec x-\vec
z)^2}\right)^d=\frac{\pi^{d/2}\Gamma(d/2)}{\Gamma(d)}\delta^d(\vec
x-\vec z) \,.
\end{eqnarray}
This logarithmic factor implies a shift of conformal dimension
of the dual operator,
\begin{eqnarray}
  \label{eq:ChangeDim}
\delta \Delta_{O_c}= h q^2 \,,\qquad\quad \d\D_{\Psi_c}  =h (q-1)^2\,.
\end{eqnarray}
We have included the shift for the fermion partner of $O_c$ which is computed  by similar steps.  We
emphasize  again that shifts of conformal dimension of bosonic and fermionic components of a supermultiplet differ because they have
different R-charges. The shifts of conformal dimension and the consequent mass shifts definitely break SUSY!

It was crucial in this calculation that the initial bulk-to-boundary
propagator is converted to a bulk derivative $\pa_\mu K$ by partial integration of the $\partial_i$ derivatives in the SUSY breaking action (\ref{eq:SUSYBreakHigh}). Then $\pa_\mu$ can be partially integrated and acts via the Ward identity on the charged field line as in the $AdS_3$ calculations in Sec. \ref{sec:R-chargedMass}.
The final answer comes from boundary terms in which the gauge fields are pinned at the location of the external R-charged particles.

The calculation of the deformation of  the 2-point function
$\<O_mO_m\>$ for R-neutral particles proceeds in a similar fashion
to that in Sec. \ref{sec:R-neutralMass}.  The $U(1)$ Ward identity
guarantees that the contribution from each loop containing R-charged
particles cancels exactly, and there are no residual boundary terms.
The $U(1)$ Ward identity is a general and exact principle, so there
are no mass corrections for R-neutral particles at any loop order.

\subsubsection{The same calculation in the CFT }\label{sec:CFTside}
It is instructive (and even easier!) to show that the same result
for the mass shift can be obtained using conformal perturbation
theory and the $U(1)$ Ward identity
\be \label{wardcft} \pa_i\<
...J^i(\vec w)O_c(\vec z)...\> \approx_{ \vec w\approx \vec z}
q\d^{(d)}(\vec w- \vec z)\<... O_c(\vec z)...\>\,.
\ee

To obtain the leading order $h$ contribution,  we insert the CFT
dual SUSY breaking operator in $\langle O_c^\dag(\vec y) O_c(\vec
z)\rangle$ and proceed as follows:
\begin{eqnarray}
  \label{eq:2ptCFT}
\delta\langle O_c^\dag(\vec y) O_c(\vec z)\rangle &\sim& h  \int d^d
\vec w_1 d^d \vec w_2 \ \langle O_c^\dag(\vec y) J_i(\vec
w_1)\frac{J_{ij}(\vec w_1-\vec w_2)}{(\vec w_1-\vec w_2)^2}J_j(\vec
w_2) O_c(\vec z)\rangle\nonumber\\
&\sim& h  \int d^d \vec w_1 d^d \vec w_2 \
\partial_i \partial_j\textrm{Log}[\frac{(\vec w_1-\vec w_2)^2}{a^2}]\langle O_c^\dag(\vec y)
J_i(\vec w_1)J_j(\vec w_2) O_c(\vec z)\rangle \,.
\end{eqnarray}
We may integrate by parts and use the Ward identity
\begin{equation}
  \label{eq:2ptCFTWI}
\partial_i \partial_j\langle
O_c^\dag(\vec y) J_i(\vec w_1)J_j(\vec w_2) O_c(\vec z)\rangle \sim
h q^2 \left[\delta^d(\vec y-\vec w_1)-\delta^d(\vec z-\vec
w_1)\right]\left[ \delta^d(\vec y-\vec w_2)-\delta^d(\vec z-\vec
w_2)\right] \langle O_c^\dag(\vec y) O_c(\vec z)\rangle \,.
\end{equation}
The integrals over $\vec w_1$ and $\vec w_2$ become trivial with the
same  result as
(\ref{eq:ChargedLeadingHighFinal}), which induces the shift of
conformal dimension  (\ref{eq:ChangeDim}).

\subsubsection{The order $h$ mass shift is exact}\label{sec:Exact}
Possible higher order contributions come from the insertion of higher powers of $S_{AdS_{d+1}}$ or $S_{CFT_d}$ in correlation functions.  To study these we carry out the explicit calculation at order $h^2$
order from the CFT viewpoint. The $h^2$ order correction to
$\delta\langle O_c^\dag(\vec y) O_c(\vec z)\rangle$ can be written
as
\begin{equation}
  \label{eq:NecklaceDiagram}
\delta_{h^2}\langle O_c^\dag(\vec y) O_c(\vec z)\rangle \sim
h^2\int d^d \vec w_1 ... d^d \vec w_4 \langle O_c^\dag(\vec y)
J_i(\vec w_1)\frac{J_{ij}(\vec w_1-\vec w_2)}{(\vec w_1-\vec
w_2)^2}J_j(\vec w_2)  J_k(\vec w_3)\frac{J_{kl}(\vec w_3-\vec
w_4)}{(\vec w_3-\vec
w_4)^2}J_l(\vec w_4) O_c(\vec z)\rangle\,.
\end{equation}
Applying conformal perturbation theory,  one sees that there are two inequivalent sets of
connected contributions which play rather different roles:

i. order $q^2$, with a Wick contraction of any pair of currents such
as \be \<\ldots J_j(\vec w_2)J_k(\vec w_3)\ldots\>\sim \<\ldots
\frac{J_{jk}(\vec w_2-\vec w_3)}{(\vec w_2-\vec
w_3)^{2(d-1)}}\ldots\>\,. \ee

ii. order $q^4$, in which one uses
(\ref{eq:DoubleLog}) and partial integration of $\pa_i\pa_j $
followed by the  $\pa_iJ^i(\vec w_1) O_c^\dagger(\vec y)$ and
$\pa_jJ^j(\vec w_2)O_c(\vec z)$ Ward identities, and then  similar
manipulations on the currents $J_k(\vec w_3)$ and $J_l(\vec w_4)$.

{\bf Order $q^2$:}  Partial integration of $\pa_j$ leads to $\<...\pa_jJ_j(\vec w_2)  J_k(\vec w_3)...\>$.  The current is conserved, but the question is whether there is a contact term in the OPE when $\vec w_2 \approx\vec w_3$.  A  regulated calculation\footnote{ One natural method, based on differential regularization,  is used in Appendix A.} shows that
\begin{eqnarray}
  \label{eq:NecklaceIdentityd2}
\partial_j \langle J^j(\vec w_2)J_k(\vec w_3)\rangle &=&
-\pi\frac{\pa}{\partial w_{3}^k}
\delta^2(\vec w_2-\vec w_3), \ \ \ \ \ \ d=2\nonumber\\
\partial_j \langle J_j(\vec w_2)J_k(\vec w_3)\rangle &=&0\,. \ \ \ \ \ \ \ \ \ \ \ \ \ \ \ \ \ \ \ \ \ \ \ \ \ \ \ \ \ \ d>2 \,.
\end{eqnarray}
There is a natural contact term in $d=2$, but none in higher dimensions. Therefore the shift
$\d_{h^2q^2}\D_{O_c}$ (and higher order shifts) vanishes for
$d\ge3$.

{\bf Order $q^4$:}  The operations described above together with
similar operations with permutations of the four currents lead to
the result \be \label{dh2q4} \delta_{h^2} \<O_c^\dag(\vec y)
O_c(\vec z)\> = -\frac12 h^2 q^4 \log^2[\frac{(\vec y-\vec
z)^2}{a^2}]\frac{1}{(\vec y - \vec z)^{2\D}}\,. \ee When added to
the order $h^0$ and $h$ terms, we find the beginning of the
exponential series that expresses the fully corrected $\<O_c^\dagger
O_c\>$ correlation in the power law form
\be \label{ococexact}
\langle O_c^{\dagger}(\vec y) O_c(\vec z)\rangle_{\rm exact}
=c(\D+hq^2)\frac{1}{(\vec y - \vec z)^{2(\D+hq^2)}}
\ee
required
by conformal symmetry.  To see the full exponential series one must
consider $n$ insertions of $S_{CFT_d}$.  For $d\ge 3$ the only
non-vanishing contributions are of order $h^n q^{2n}$ and come from
$n$ applications of (\ref{eq:DoubleLog}) with subsequent partial
integration and use of (\ref{wardcft}). The combinatoric structure
is the same as that of $AdS_3$ scenario \cite{Dong:2014tsa} and
leads to exponentiation.

\subsubsection{Exact marginality}\label{sec:marginality}
As stated in Sec. \ref{sec:SUSYAdS},  insertions of $S_{AdS_{d+1}}$
in any correlator are expected to preserve its conformal structure.
Thus  $S_{AdS_{d+1}}$ is effectively exactly marginal.  The result
(\ref{ococexact}) above is one indication that this is true.

We may also verify this by calculating the correction to $\<J_i(\vec y)J_j(\vec z)\>$ which at the leading order is given by
\begin{eqnarray}
  \label{eq:JDim}
\delta_h\langle J_a(\vec y) J_b(\vec z) \rangle &\sim& h\int d^d\vec
w_1 d^d\vec w_2\ \langle J_a(\vec y) J_i(\vec w_1) \frac{J_{ij}(\vec
w_1-\vec w_2)}{(\vec w_1-\vec w_2)^2}J_j(\vec w_2) J_b(\vec z)
\rangle\nonumber\\
&\sim& h\int d^d\vec w_1 d^d\vec w_2\ \langle J_a(\vec y) J_i(\vec
w_1) \partial_i\partial_j\textrm{Log}[(\vec w_1-\vec w_2)^2]J_j(\vec
w_2) J_b(\vec z) \rangle \,.
\end{eqnarray}
Using integration by parts and  (\ref{eq:NecklaceIdentityd2}), we
find $\delta_h\langle J_a(\vec y) J_b(\vec z) \rangle =0$ for $d>2$.
Higher order corrections also vanish,  and the conformal properties
are preserved.\footnote{For $d=2$ the right hand side of
(\ref{eq:NecklaceIdentityd2}) does not vanish, and there is a
residual correction proportional to the undeformed correlator
$1/(\vec y-\vec z)^2$.  The conformal structure is preserved in a
less trivial fashion.}

\subsubsection{Gravity mediated SUSY breaking effects vanish}\label{sec:gravity}
\begin{figure}[h]
\begin{center}
\includegraphics[width=0.2\textwidth]{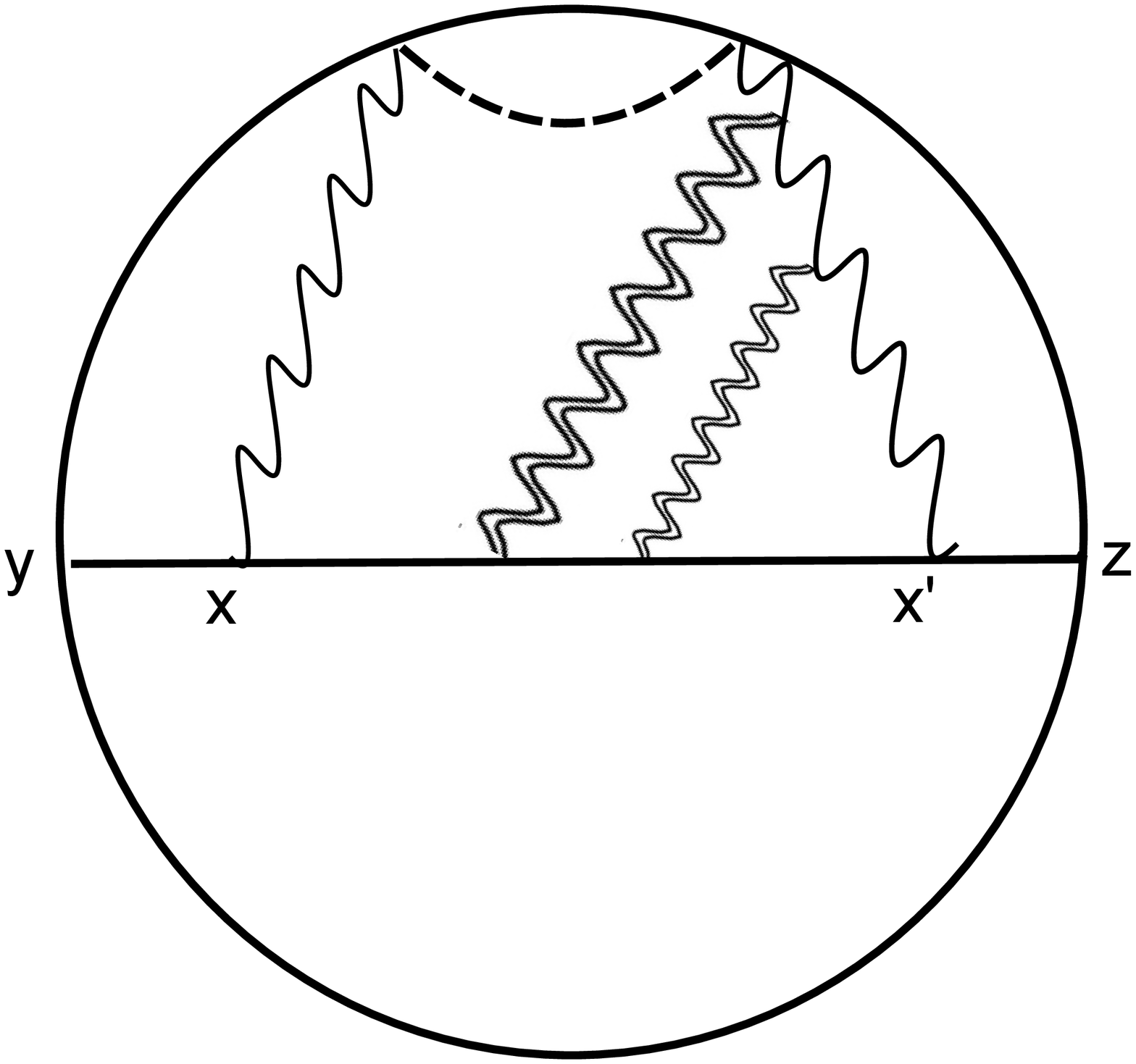}
\hspace*{0.35cm}
\includegraphics[width=0.205\textwidth]{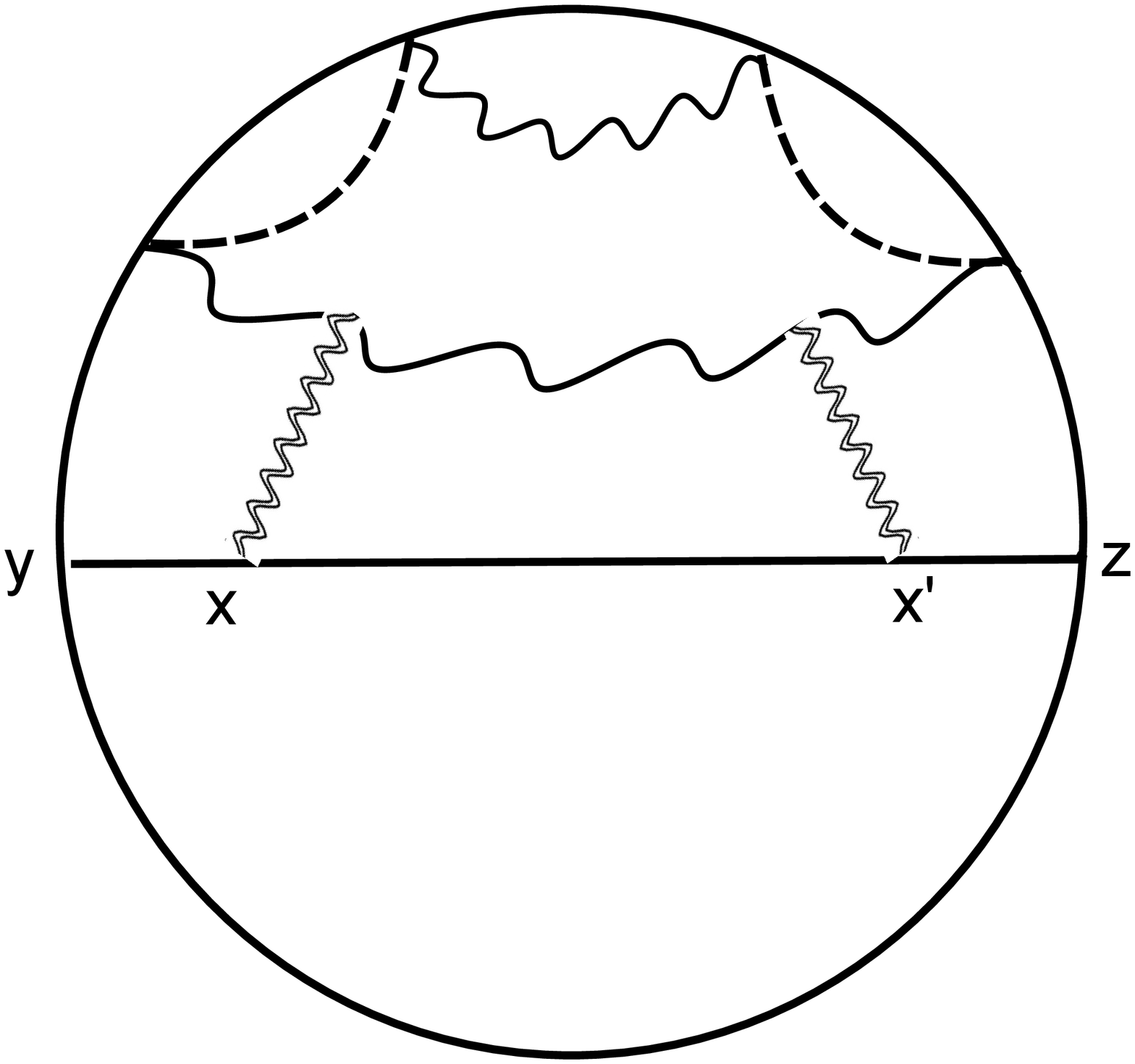}
\hspace*{0.35cm}
\includegraphics[width=0.2\textwidth]{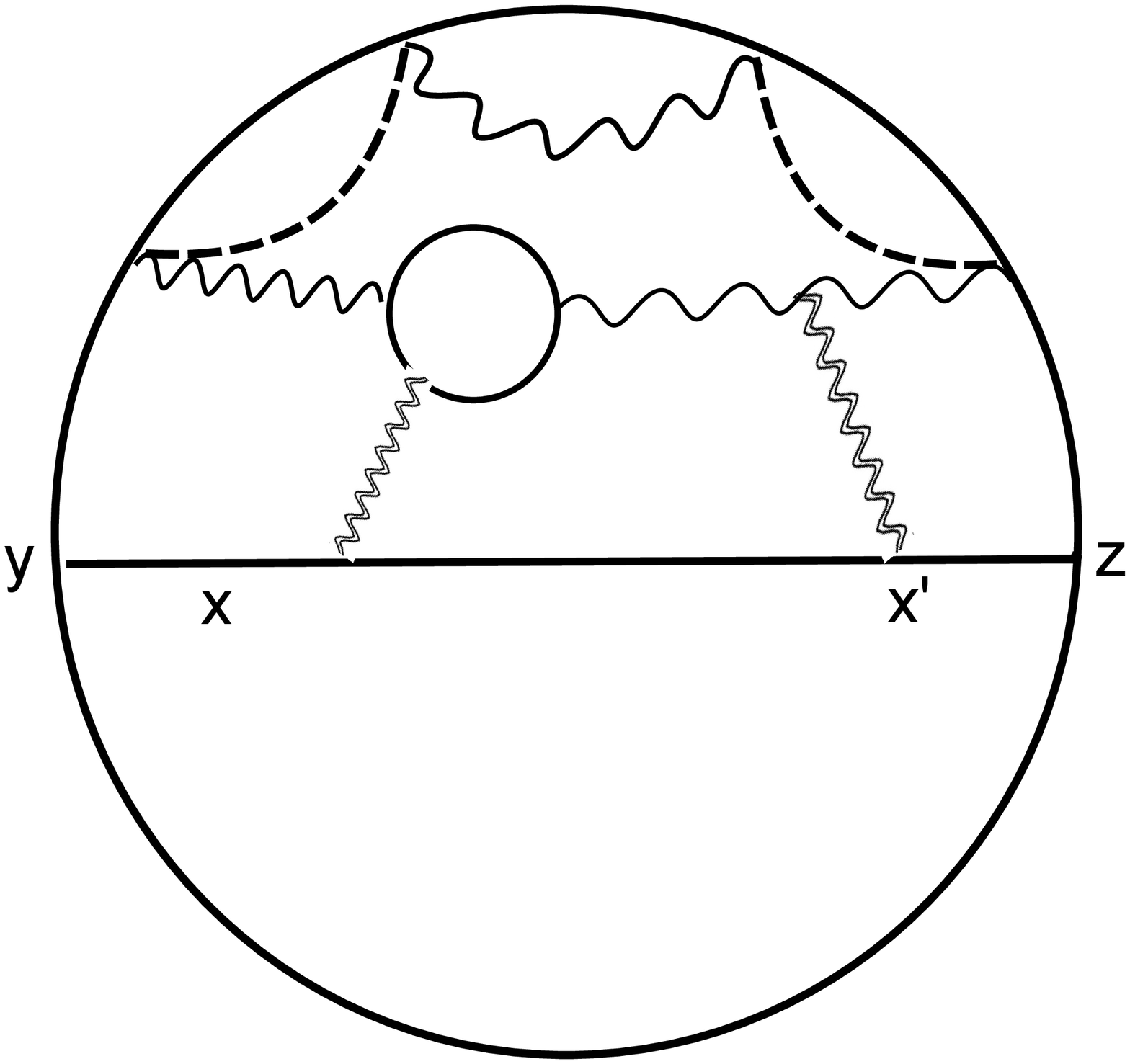}
\caption{Witten diagrams which involve gravitons. The double lines indicate gravitons. The first diagram describes the SUSY breaking corrections,
mediated by the graviton, to R-charged particles. The horizontal line in the second and third diagrams can be charged or neutral.
 In the  third diagram a graviton couples to a charged loop.}
\label{fig:graviton}
\end{center}
\end{figure}

As discussed in the previous section, SUSY is explicitly broken in our theory by an effectively marginal deformation.
Unlike common  SUSY breaking mechanisms, especially spontaneous breaking,
 there is no energy scale above which SUSY is restored.
Then one  concern in our SUSY breaking mechanism is whether
the diagrams involving graviton loops can introduce additional SUSY
breaking corrections to R-charged particles. More seriously, if SUSY
breaking effects can propagate to R-neutral particles through
graviton loops, the SUSY breaking effects may be huge due to the
absence of SUSY restoring scale. These concerns can be explored by
studying the Witten diagrams shown in Fig. \ref{fig:graviton}.

There are two categories of diagrams involving the bulk graviton.

1. The graviton
line ends on an R-gauge photon emitted directly from the SUSY breaking vertex.

2. The graviton line ends on an internal loop of an R-charged matter field that also
couples to an R-gauge field from the SUSY breaking vertex.

The main new observation needed to show that the first class of diagrams vanish
is that the graviton couples to $A_\m$
 through the field strength
$F_{\m\n}$, i.e.
\begin{eqnarray}
  \label{eq:GTcoupling}
L\sim \d g_{\mu\nu} T^{\mu\nu}&\sim& \d g_{\mu\nu}
(F^{\mu\rho}F^\nu{}_\rho-\frac{1}{4}g^{\mu\nu}F^2) \,.
\end{eqnarray}
If the R-gauge boson comes directly from the SUSY
breaking vertex, its bulk boundary propagator couples to $F_{\m\n}$ as $\pa_{[\m} K_{\n]}^i$. This vanishes after partial integration
 on the boundary which converts the photon propagator to a bulk
total derivative,  i.e. $\pa_i \pa_{[\m} K_{\n]}^i= \pa_{[\m}\pa_{\n]} K =0$.

Diagrams containing charged loops coupled both to a graviton and one
or more  gauge fields from the SUSY breaking vertex vanish by the
same arguments used in Sec. \ref{sec:SUSYAdSCorrectionRcharged}. The
presence of a graviton  does not affect the validity of the $U(1)$
Ward identity.

\section{Flat spacetime limit?}\label{sec:flat}
 \setcounter{equation}{0}
The approach to the little hierarchy problem described in this paper works in anti-de Sitter spacetime, but it would be far more interesting if SUSY breaking effects survive in the limit $L\to\infty$  to Minkowski space.
We give a preliminary discussion of this limit;  details are under investigation.

At large $L$, the AdS/CFT mass formulas for the superpartners $\phi_c$ and $\Psi_c$ (which we now denote by B and F, respectively) behave as
\bea
\D_B &=& \frac{d}{2} + \sqrt{\frac{d^2}{4} + m_B^2L^2} \to m_BL \,,\\
\D_F &=& \frac{d}{2} +|m_FL|\to m_FL \,.
\eea
The SUSY breaking shifts of (\ref{eq:ChangeDim}) are exact for any value of the coupling $h$.  For large $h$ the boson-fermion mass splitting is
\be
m_B-m_F = \frac{\D_B-\D_F}{L} = \frac {h(2q-1)}{L}\,.
\ee
The mass splitting survives in the flat spacetime limit if we write $h=m_0L$, where $m_0$ is a chosen mass scale, and scale $L\to \infty$.  In the limit we find
\be
m_B-m_F =(2q-1)m_0\,.
\ee

We now bring into the discussion the Mellin representation of
AdS/CFT correlators proposed by Mack \cite{Mack:2009mi,Mack:2009gy}
and further developed by Penedones,  Fitzpatrick, Kaplan, Raju, and
van Rees
\cite{Penedones:2010ue,Fitzpatrick:2011ia,Fitzpatrick:2011dm}.  The
current status of this interesting research program was discussed by
Penedones at the Strings 2015 conference \cite{Penedones2015}.  In
the limit $L\to \infty$, the Mellin representation yields a quite
direct relation between  $n$-point correlation functions computed in
a given bulk theory in $AdS_{d+1}$ and the flat space $S$-matrix of
that theory.  It may be  the case that the relation can be extended
to include the SUSY breaking effects generated by the boundary
interaction of Sec. \ref{sec:SUSYAdS} above.

\section{Discussion}\lab{Disc}
In this paper, we propose a new SUSY breaking mechanism in
$AdS_{d+1},~d\ge3$, in which breaking effects are generated by a
boundary action.  This is a generalization of the mechanism
studied in $AdS_3$. The most appealing fact of this new SUSY
breaking mechanism based on gauged $U(1)_R$ symmetry is that
breaking effects of arbitrary strength can be evaluated exactly. One
can  thereby achieve large mass splitting between boson and fermion
partners in a supermultiplet, while $R$-neutral fields do not
receive SUSY breaking corrections to any loop order. Another
important feature of our model is that diagrams involving graviton loops do not introduce any further SUSY breaking corrections.

The SUSY breaking action preserves the conformal properties of AdS/CFT correlation functions and is therefore effectively exactly marginal. Most effects can be calculated by two methods.

i. We can evaluate SUSY breaking in Witten diagrams exactly because the bulk-to-boundary propagator of the R-gauge boson is converted into a total derivative at the bulk point where it couples to R-charged fields. Partial integration and use of the $U(1)$ Ward identity permit straightforward integration over each bulk insertion point, and the R-gauge bosons become pinned at the boundary. The remaining boundary integrals can also be done.

ii. We can evaluate the same effect in the boundary CFT using
conformal perturbation theory and results always agree.

So far our mechanism operates in $AdS$ spacetime, but there are
several directions to pursue toward possible  phenomenological
models in flat spacetime. As suggested in Sec. \ref{sec:flat}, it may
be possible to extend  the recently developed connection between the
Mellin transform of AdS/CFT correlators and the flat space S-matrix
to include our SUSY breaking mechanism. This can be pursued in the flat space limit of $AdS_4$.  Alternatively in $AdS_5$ one can consider the implementation of boundary SUSY breaking in the supersymmetric Randall-Sundrum
model \cite{Randall:1999ee,Gherghetta:2000qt,Gherghetta:2011wc}. Or
equivalently, one can study the phenomenology of new physics generated by deformations of a four-dimensional strongly coupled SCFT.  We leave these interesting topics to future work.

\section*{Acknowledgments}  We thank Nima Arkani-Hamed, Nathaniel Craig, Savas Dimopoulos, Henriette Elvang,  Liam Fitzpatrick, Guy Gur-Ari,  Don Marolf, Joao Penedones, Nati Seiberg, Steve Shenker, David Simmons-Duffin, and Eva Silverstein for useful discussions and correspondence.  XD is supported by the National Science Foundation
under grant PHY-1314311 and by a Zurich Financial Services
Membership at the Institute for Advanced Study. The research of DZF
is supported  by NSF grant PHY-0967299 and DOE Grant No.
DE-SC0012567. YZ is supported by DE- SC0007859.

\appendix

\section{Differential Regularization of $\<J_i(\vec w)J_j(\vec w')\>\sim \frac{J_{ij}(\vec w -\vec w')}{(\vec w -\vec w')^{2(d-1)}}$}
\setcounter{equation}{0}

We apply differential regularization \cite{Freedman:1991tk} to argue that there is no contact term on the right side of (\ref{eq:NecklaceIdentityd2}) for boundary dimension $d\ge 3$. The basic idea of this method is quite simple.  A singular function such as $1/(\vec w -\vec w')^{2p}$  in an integrated expression can usually be expressed as  the total derivative of a less singular function.  For example, one can use the identity (for $\vec w-\vec w'  \ne 0$ and $2p -d >0$)
\be \lab{ident1}
\frac{1}{(\vec w -\vec w')^{2p}} = \frac{1}{2(p-1)(2p-d)}\Box\frac{1}{(\vec w -\vec w')^{2(p-1)}}\,.
\ee
If the derivatives in $\Box$ are integrated by parts and applied to other factors in the amplitude under study, the  degree of divergence  of the initial integral is decreased. In the special case $2p=d$ of a $\log$ divergence we use instead
\be\lab{ident2}
\frac{1}{|\vec w|^d} = -\frac{1}{2d-4}\Box\frac{\ln (\vec w^2M^2)}{|\vec w|^{d-2}}\,.
\ee
The integration constant $M^2$ can often be interpreted as the renormalization group scale.

In our application to the tensor  $\frac{J_{ij}(\vec w)}{|\vec w |^{2(d-1)}} =\frac{\d_{ij} - 2w_iw_j/\vec w^2}{|\vec w |^{2(d-1)}} $, we  need another identity
\be\lab{ident3}
\frac{w_iw_j}{|\vec w|^{2d}} =\frac{1}{4(d-1)(d-2)}\bigg[ \pa_i\pa_j\frac{1}{|\vec w|^{2(d-2)}}-\frac{1}{d-2}\Box\frac{1}{|\vec w|^{2(d-2)}}\bigg] \,.
\ee
We  used (\ref{ident1}) with $2p =2(d-1)$ above and now again to obtain
\be\lab{ident4}
\frac{J_{ij}(\vec w)}{|\vec w |^{2(d-1)}}= \frac{1}{2(d-1)(d-2)}[\d_{ij}\Box -\pa_i\pa_j]\frac{1}{|\vec w|^{2(d-2)}}\,.
\ee
It is no surprise that the result is proportional to the transverse projector $\d_{ij}\Box -\pa_i\pa_j$ since the $\<J_iJ_j\>$ correlator is formally conserved.
This result  is valid for $d\ge3.$

For $d=2$, current correlator is not manifestly conserved.  However,
the identity (\ref{eq:DoubleLog})  provides a regularization and gives the well known contact term already noted in (\ref{eq:recursionAdS3}).  The correlator
is conserved for separated points.

For $d=3$, the identity (\ref{ident4}) provides a regularization, since the final power $1/|\vec w|^2$ is integrable.
To see how the identity is used,  we return to (\ref{eq:NecklaceDiagram}) and substitute the regulated expression for the operator product $J_j(\vec w_2) J_k(\vec w_3)$ and substitute  (\ref{eq:DoubleLog})  for the factor $J_{ij}(\vec w_1-\vec w_2)/(\vec w_1-\vec w_2)^2$. We then partially integrate the projector to obtain,  in an entirely finite manner,
\be \lab{last}
(\delta_{jk} \Box -\pa_j\pa_k) \pa_i\pa_j \log((\vec w_1-\vec w_2)^2) =0\,.
\ee
For $d=4$,  we need to use (\ref{ident2}) to regulate the operator product completely.  After further partial integration we still find (\ref{last}).  This procedure may be applied for $d\ge 5$ after further use of (\ref{ident1}) and/or (\ref{ident2}) to regulate the singular factor $1/(\vec w_2-\vec w_3)^{2(d-2)}$ in (\ref{ident4}).
In this way we see that there are no contact terms in (\ref{eq:NecklaceIdentityd2}) for $d\ge 3$.


\end{document}